\documentclass[11pt]{article}

\usepackage[preprint]{acl}

\usepackage{times}
\usepackage{latexsym}

\usepackage[T1]{fontenc}

\usepackage[utf8]{inputenc}

\usepackage{microtype}

\usepackage{inconsolata}

\usepackage{algorithm}
\usepackage{algpseudocode}
\usepackage{amsmath}
\usepackage{tabularx}
\usepackage{graphicx}
\usepackage{pifont}
\usepackage{multirow}
\usepackage{listings}
\usepackage{makecell}
\usepackage{utfsym}
\usepackage{amssymb}

\title{SC-MAS: Constructing Cost-Efficient Multi-Agent Systems with Edge-Level Heterogeneous Collaboration}

\author{
  Di Zhao$^{1}$,
  Longhui Ma$^{1}$,
  Siwei Wang$^{2*}$,
  Miao Wang$^{2*}$,
  Yi Kong$^{1}$
  \\
  $^{1}$College of Computer Science and Technology, National University of Defense Technology
  \\
  $^{2}$Academy of Military Sciences
  \\
  \small{
      {zhaodi@nudt.edu.cn}
  }
}

\begin{document}
\maketitle

{\renewcommand{\thefootnote}{}
\footnotetext{*Corresponding authors}}

\begin{abstract}

Large Language Model (LLM)-based Multi-Agent Systems (MAS) enhance complex problem solving through multi-agent collaboration, but often incur substantially higher costs than single-agent systems.
Recent MAS routing methods aim to balance performance and overhead by dynamically selecting agent roles and language models. 
However, these approaches typically rely on a \emph{homogeneous collaboration mode}, where all agents follow the same interaction pattern, limiting collaboration flexibility across different roles.
Motivated by \emph{Social Capital Theory}, which emphasizes that different roles benefit from distinct forms of collaboration, we propose \textbf{SC-MAS}, a framework for constructing \emph{heterogeneous and cost-efficient} multi-agent systems.
\textbf{SC-MAS} models MAS as directed graphs, where edges explicitly represent pairwise collaboration strategies, allowing different agent pairs to interact through tailored communication patterns. 
Given an input query, a unified controller progressively constructs an executable MAS by selecting task-relevant agent roles, assigning edge-level collaboration strategies, and allocating appropriate LLM backbones to individual agents.
Experiments on multiple benchmarks demonstrate the effectiveness of \textbf{SC-MAS}. In particular, \textbf{SC-MAS} improves accuracy by 3.35\% on MMLU while reducing inference cost by 15.38\%, and achieves a 3.53\% accuracy gain with a 12.13\% cost reduction on MBPP. 
These results validate the feasibility of \textbf{SC-MAS} and highlight the effectiveness of heterogeneous collaboration in multi-agent systems.

\end{abstract}

\section{Introduction}
\label{sec::introduction}
In recent years, Large Language Model (LLM)-based agents have achieved remarkable success across a wide range of tasks, including software development \cite{hong2023metagpt, chan2023chateval, li2023camel}, social simulation \cite{park2023generative, gao2023s}, gaming \cite{akata2023playing, wang2023voyager, tan2024towards}.
Building on the impressive capabilities of single agents, LLM-based Multi-Agent Systems (MAS) have emerged as a powerful paradigm for solving complex problems by leveraging collaboration among agents with specialized roles\cite{guo2024large, qian-etal-2024-chatdev}.
Meanwhile, the LLM ecosystem has become increasingly diverse, ranging from lightweight and low-cost models to large-scale, high-performance yet expensive alternatives \cite{matarazzo2025surveylargelanguagemodels, chen2024frugalgpt, ong2025routellm}.
This diversity complicates model selection, as larger models do not consistently outperform smaller ones across all tasks and domains \cite{abdin2024phi3technicalreporthighly, lepagnol-etal-2024-small, shen-etal-2024-small}.
Consequently, determining how to effectively allocate LLM resources under performance and cost constraints has become a central challenge in practical LLM-based systems.

To address this challenge, prior work \cite{chen2024frugalgpt, hu2024routerbench} has extensively studied \textit{LLM routing}, which aims to dynamically select appropriate language models for a given input.
Early approaches relied on auxiliary encoders such as BERT \cite{devlin-etal-2019-bert} to decide whether to invoke stronger models \cite{chen2024frugalgpt, ding2024hybrid, ong2025routellm}, while more recent methods formulated routing as an optimization problem over performance-cost trade-offs \cite{feng2025graphrouter, mohammadshahi2024routoolearningroutelarge, dai2024costeffectiveonlinemultillmselection}.
Despite their effectiveness, these approaches primarily focus on single-agent settings and do not consider collaborative interactions among multiple agents. \cite{chen2024frugalgpt, ong2025routellm, feng2025graphrouter}.
In parallel, substantial progress has been made in designing dynamic multi-agent systems \cite{li2024more, chen2024are, wang2025mixtureofagents}.
Several studies model MAS as directed acyclic graphs (DAGs) and learn query-specific structures to improve performance \cite{gptswarm2024, wang2025agentdropout, zhang2025cut}.
However, these methods typically assume homogeneous agent behaviors or rely on a single LLM backbone, leaving the cost dimension largely unexplored.
Conversely, naively applying single-agent LLM routing independently to each agent ignores the rich interaction patterns that emerge from agent collaboration.
These limitations highlight the need for a unified framework that jointly considers \emph{which agents to include}, \emph{how they collaborate}, and \emph{which LLM resources they should employ}.

Human collaboration provides an important perspective on this problem.
According to \emph{Social Capital Theory} \cite{coleman1988social, putnam2001social}, effective collective behavior arises not only from individual capabilities but also from the relational ties between individuals.
In practice, different pairs of collaborators often adopt distinct interaction patterns, such as critique, debate, depending on their roles and shared objectives.
Motivated by this insight, we model a multi-agent system as a structured network, where agents correspond to nodes and edges explicitly encode pairwise collaboration strategies.

Based on this perspective, we propose \textbf{SC-MAS}, a \textbf{S}ocial \textbf{C}apital-driven framework for constructing heterogeneous and cost-efficient \textbf{M}ulti-\textbf{A}gent \textbf{S}ystems.
Given an input query, \textbf{SC-MAS} progressively constructs an executable MAS through three stages:
(i) selecting task-relevant agent roles from a candidate pool,
(ii) establishing edge-level collaboration strategies that explicitly model pairwise interactions between agents, and 
(iii) assigning appropriate LLM backbones to each agent based on both their roles and collaborative context.
These components are jointly optimized to balance task performance and computational cost, enabling flexible and efficient multi-agent collaboration.

Our contributions can be summarized as follows:
\begin{itemize}
\item We introduce a social capital-driven perspective on multi-agent system construction, emphasizing the role of relational structures and heterogeneous collaboration strategies in effective MAS design.
\item We propose \textbf{SC-MAS}, a modular framework that integrates agent selection, collaboration structure construction, and LLM assignment into a unified and efficient system.
\item Experiments on five benchmarks show that \textbf{SC-MAS} reduces token consumption by 11.17\% to 16.35\% while improving accuracy by 1.46\% to 3.34\% over state-of-the-art methods, validating both its effectiveness and efficiency.
\end{itemize}

\section{Related Work}
\label{sec::related-work}

\paragraph{Dynamic Multi-Agent Systems} Dynamic MAS aim to adapt agent compositions or interaction structures according to task requirements.
Existing work has explored search-based methods, such as Monte Carlo Tree Search \cite{hu2025automated, zhang2025evoflowevolvingdiverseagentic, shang2025agentsquare} and evolutionary algorithms \cite{zhang2025cut}, to discover effective agent structures.
Recent approaches, including DyLAN \cite{liu2024dynamic}, GPTSwarm \cite{gptswarm2024}, and AgentPrune \cite{zhang2025cut}, model MAS as directed acyclic graphs and optimize collaboration structures in a query-specific manner.
However, these methods typically rely on a single LLM backbone and incur substantial computational overhead, limiting their ability to jointly optimize collaboration diversity and cost efficiency in heterogeneous LLM environments.

\paragraph{Single-Agent LLM Routing} LLM routing has been extensively studied in single-agent settings as a means to balance performance and inference cost.
Early work focused on binary routing between small and large models \cite{chen2024frugalgpt, ding2024hybrid, ong2025routellm}, while more recent approaches extended routing to multi-model selection and optimization frameworks \cite{dai2025costeffective, feng2025graphrouter, chen2024routerdc}.
Although effective for single-agent systems, these methods do not account for collaborative interactions among multiple agents and therefore cannot be directly applied to MAS.

\paragraph{Multi-Agent System Routing} More recently, MasRouter \cite{yue2025masrouter} introduced MAS routing by jointly determining agent roles, collaboration modes, and LLM allocation.
While MasRouter demonstrates strong performance, it adopts a \emph{graph-level} collaboration strategy, where all agents share the same interaction pattern.
Such a homogeneous design limits the expressiveness of multi-agent collaboration and fails to capture fine-grained, pairwise interaction patterns that naturally arise in complex tasks.
In contrast, our work focuses on \emph{edge-level} collaboration modeling, enabling heterogeneous interaction strategies between different agent pairs.
By explicitly representing collaboration strategies as edges in a graph, \textbf{SC-MAS} provides a more expressive and flexible framework for adaptive multi-agent system construction.

\section{Preliminaries}
\label{sec::preliminaries}

\subsection{MAS as graph with edge strategies}
\label{sec::define_mas}

Existing graph-based formulations of MAS \cite{gptswarm2024, qian2025scaling, wang2025agentdropout} typically represent agents as nodes and communication flows as edges. 
While effective for information passing, such formulations largely overlook the \emph{collaborative semantics} that govern how agents interact.
To explicitly capture heterogeneous collaboration patterns, we model MAS as directed graphs whose edges encode collaboration strategies rather than mere message transmission.
Formally, we represent a MAS as a directed acyclic graph $G= (V, E, L)$. 
The graph consists of agent nodes $V$, their associated LLMs $L = \{ l_v \}_{v \in V}$, and directed edges $ E\subseteq V\times V\times S $. 
Each edge $e = (u, v, s) \in E$ denotes agent $u$ employing strategy $s \in S$ with respect to agent $v$.
Strategies are categorized into \textbf{edge strategy} ($S_{edge}, \ u \ne v$) and \textbf{self-loop strategy} ($S_{self}, \ u = v$) types, allowing representation of both collaboration patterns (e.g. Debate \cite{liang-etal-2024-encouraging, du2024improving}, Chain \cite{qian2025scaling}) and internal reasoning (e.g. Chain-of-Thought \cite{cot2022}, Reflection \cite{shinn2023reflexion}).
Information is exchanged between the two agents via edge strategy $s_{eg} \in S_{edge}$; in particular, the agent uses self-loop strategy $s_{sl} \in S_{self}$ each time to give a response.

Given a query $q$, the graph $G$ executes topologically: each agent $v$ performs its inference process $f_v$ (using LLM $l_v$) to produce its own output $o_v = f_v(q, Z_v)$ based on $q$ and outputs $Z_v$ from its predecessors.
The execution of the graph is described in Algorithm~\ref{alg1}.

\begin{algorithm}[h]
\caption{Graph Execution} 
\label{alg1} 
\begin{algorithmic}[1]
\Require MAS graph $G=(V, E, L)$, query $q$.
\Ensure Final output collection $O_{final}$.
\State Initialize $Z \leftarrow \text{empty list}$ for each node in the start
\For{each node $u$ in TopologicalSort($V$)}
    \State Let $s_{sl}^u$ be the self-loop strategy for $u$ from $S_{self}$ (if $u \to u$ edge exists, else null)
    \For{each edge $(u, v, s_{eg})$ in $E$ where $u \ne v$ and source is $u$}
        \While{edge strategy $s_{eg}$ has not been executed}
            \State Let $s_{sl}^v$ be the self-loop strategy for $v$ from $S_{self}$ (if $v \to v$ edge exists, else null)

            \Comment{agent $u$ and $v$ continuously exchange information through the strategy $s_{eg}$}
            \State $Z_v \leftarrow Z_v\cup f_u(q, Z_u, s_{sl}^u, s_{eg})$
            \State $Z_u \leftarrow Z_u\cup f_v(q, Z_v, s_{sl}^v, s_{eg})$
        \EndWhile
    \EndFor
    \If{$u$ has no outgoing edges $(u, v, s_{eg})$ where $u \ne v$}

        \State $O_{final} \leftarrow O_{final} \cup f_u(q, Z_u, s_{sl}^u) $
    \EndIf
\EndFor
\end{algorithmic} 
\end{algorithm}

\subsection{Search Space}
\label{sec::search-space}

The search space for MAS configurations is defined over sets of predefined agent roles $\mathbb{V} = \{ v_i \}_{i=1}^{N_V}$, available LLM backbones $\mathbb{L} = \{ l_i \}_{i=1}^{N_L}$, edge strategies $\mathbb{S}_{edge} = \{ s_i \}_{i=1}^{N_{eg}}$, and self-loop strategies $\mathbb{S}_{self} = \{ s_i \}_{i=1}^{N_{sl}}$. We define the search space components $(\mathbb{V}, \mathbb{S}_{edge}, \mathbb{S}_{self}, \mathbb{L})$ as $\mathbb{G}$.

Constructing an instance $\mathcal{G} = (\mathcal{V}, \mathcal{E}, \mathcal{L}) \in \mathbb{G}$ requires selecting $d$ agents $\mathcal{V} \subseteq \mathbb{V}$, assigning LLMs $\mathcal{L} \subseteq \mathbb{L}$, and choosing an edge set $\mathcal{E}$.
Potential edges $(u, v, s)$ connect $u, v \in \mathcal{V}$ using strategy $s \in \mathbb{S}_{edge}$ (for $u \ne v$) or $s \in \mathbb{S}_{self}$ (for $u = v$).
For a chosen $\mathcal{V}$, there are $(N_{sl} \cdot N_{eg}^{d-1})^d$ potential edge sets defined as $\mathbb{E}$.
A critical constraint mandates that the resulting inter-agent graph structure must be a DAG for valid execution using Algorithm~\ref{alg1}.
Self-loop strategies, representing local node execution policies, do not affect this topological constraint. 
The DAG requirement significantly prunes the space of valid edge configurations.

\subsection{Query-conditioned MAS Construction}
\label{sec::define_mas_construction}

MAS construction is defined as the problem of learning a mapping from an input query $q$ to an optimal, executable MAS graph $\mathcal{G} = (\mathcal{V}, \mathcal{E}, \mathcal{L})$.
This mapping is stochastic, represented by a conditional probability distribution $\mathbb{P}(\mathcal{G} \mid q)$ over valid graphs constructible from the search space components $(\mathbb{V}, \mathbb{S}_{edge}, \mathbb{S}_{self}, \mathbb{L})$ and satisfying the DAG constraint on inter-agent edges.
The mapping is formally defined as:
\begin{equation}
\begin{gathered}
    f: \mathbb{V} \times \mathbb{S}_{edge} \times \mathbb{S}_{self} \times \mathbb{L} \rightarrow \mathcal{G}, \\\pi(\mathcal{G}) = \mathbb{P}(\{ \mathcal{V}, \mathcal{E}, \mathcal{L} \} \mid q) \\\mathcal{V} \subseteq \mathbb{V}, \mathcal{E} \subseteq \mathbb{E}, \mathcal{L} \subseteq \mathbb{L},  
    \lvert \mathcal{V} \rvert = \lvert \mathcal{L} \rvert
\end{gathered}
\end{equation}
where $\pi(\mathcal{G})$ represents the probability of selecting $\mathcal{G}$ under condition query $q$.

\subsection{Optimization Objective}
\label{sec::define_optimization_obj}

The objective is to learn the conditional distribution $\mathbb{P}(\mathcal{G} \mid q)$ that maximizes expected task utility while penalizing execution cost.
Given a dataset $\mathcal{D}$ of query-answer pairs $(q, a)$, the optimization problem is formulated as:
\begin{align}
\label{eq:objective}
    \max_{\mathbb{P}(\mathcal{G} \mid q)} 
    \mathbf{E}_{\underset{\mathcal{G} \in \mathbb{G} \sim \mathbb{P}(\mathcal{G} \mid q)}
    {(q,a) \sim \mathcal{D},}}
    \
    [
    \underbrace{U(\mathcal{G}; q, a)}_{\text{Utility}} - 
    \lambda \cdot \underbrace{C(\mathcal{G}; q)}_{\text{Cost}}
    ]
\end{align}
where $U(\mathcal{G} ; q, a)$ quantifies the utility of the graph $\mathcal{G}$ (e.g. accuracy evaluated using $a$), $C(\mathcal{G}; q)$ measures the associated execution cost (e.g., API calls, token usage), and $\lambda \ge 0$ is a hyperparameter controlling the trade-off between utility and cost.

\begin{figure*}[htb]
    \centering
    \includegraphics[width=1\linewidth]{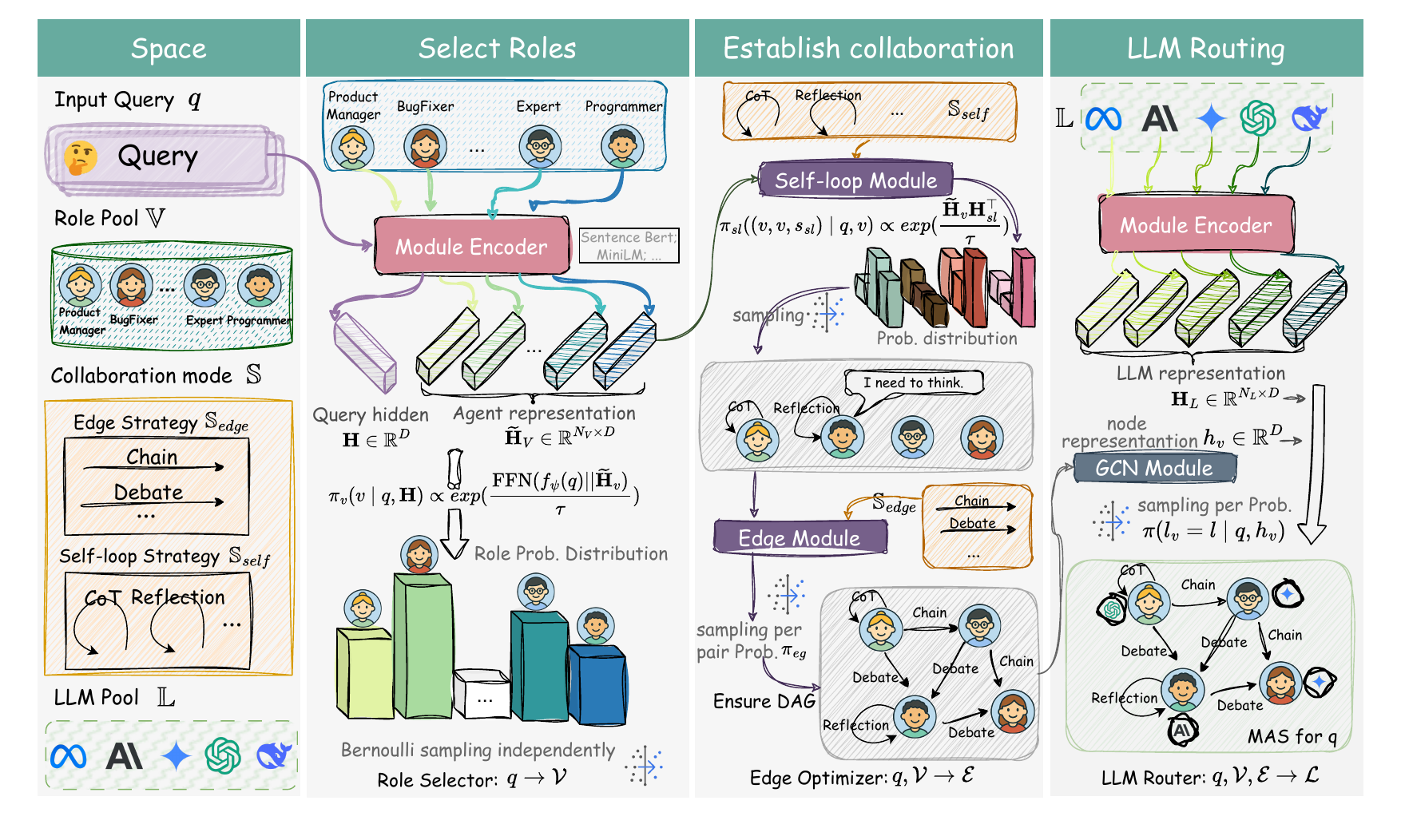}
    \caption{The overall framework of our proposed \textbf{SC-MAS}. The roles for accomplishing the query are first selected from the candidate pool, followed by establishing the collaboration relationships between these selected roles, and finally assigning the appropriate LLMs, thus generating an executable graph representation of the heterogeneous collaboration strategies MAS.}
    \label{fig::arc-main}
\end{figure*}

\section{SC-MAS}
\label{sec::SC-MAS}

Figure~\ref{fig::arc-main} depicts the \textbf{SC-MAS} architecture. 
For an input query $q$, \textbf{SC-MAS} constructs an executable MAS graph $\mathcal{G} = (\mathcal{V}, \mathcal{E}, \mathcal{L})$ via three sequential components: Node Selector ($\mathbb{F}_{\theta_v}$, Section~\ref{sec_node}), Edge Optimizer ($\mathbb{F}_{\theta_e}$, Section~\ref{sec_edge}), and LLM router ($\mathbb{F}_{\theta_l}$, Section~\ref{sec_llm}).
The overall generative process for $\mathcal{G}$ given $q$ is modeled as $\mathbb{F}_{\theta}(\mathcal{G} \mid q)$, factorized into these components: $\mathbb{F}_{\theta} = \mathbb{F}_{\theta_v} \circ \mathbb{F}_{\theta_e} \circ \mathbb{F}_{\theta_l}$.
After executing the generated graph $\mathcal{G}$ to obtain an answer $a$, the parameters $\theta = (\theta_v, \theta_e, \theta_l)$ are updated based on the utility $U(\mathcal{G}; q, a)$ and cost $C(\mathcal{G}; q)$ feedback, following the objective in Eq.~\ref{eq:objective}.

This design choice is grounded in a human-centric paradigm informed by \textit{Social Capital Theory} \cite{coleman1988social}, \cite{putnam2001social}, which posits that cooperative behavior emerges initially at the individual level. 
According to this theory, individuals form connections and engage in collaboration based on mutual trust and benefit, ultimately leading to the development of structured, goal-oriented networks.

\subsection{Node Selector}
\label{sec_node}

The Node Selector identifies the optimal subset of agents $\mathcal{V} \subseteq \mathbb{V}$ based on the query $q$.
Prior work has demonstrated that latent variable-based modeling is effective for adaptive agent selection in dynamic multi-agent systems \cite{yue2025masrouter}, motivating its use in our method.
Therefore, in order to model the complex relationship between $q$ and the potential agents $v \in \mathbb{V}$, we employ a variational latent variable model:
\begin{equation}
\begin{gathered}
\mathbb{F}_{\theta_v}(\mathcal{V} \mid q) = \int p_v(\mathcal{V} \mid \mathbf{H}) \  p_h(\mathbf{H} \mid q) \ d \mathbf{H}, 
\\ \mathcal{V} \subseteq \mathbb{V}
\end{gathered}
\end{equation}
Here, $p_h(\mathbf{H} \mid q) = \mathcal{N}(\mathbf{H}; \mu_t(q), \text{diag}(\sigma_t^2(q)))$ is a Gaussian prior over the latent variable $\mathbf{H}$, conditioned on $q$.
The generator $p_v(\mathcal{V} \mid \mathbf{H})$ models the selection of nodes using a factorized Bernoulli distribution:
\begin{equation}
\begin{gathered}
    p_v(\mathcal{V} \mid \mathbf{H}) = \prod_{v \in \mathbb{V}} \pi_v(v \mid q, \mathbf{H}) \\
    \pi_v(v \mid q, \mathbf{H}) \propto
     exp(\text{FFN}(f_\psi(q) || \widetilde{\mathbf{H}}_v) / \tau) \\ \widetilde{\mathbf{H}}_v = g_\phi(f_\psi(v), \mathbf{H}), \quad \tau > 0
\label{eq:node_prob}
\end{gathered}
\end{equation}
where $\pi_v$ is the selection probability for node $v$, computed using a feed-forward network (FFN).
$f_\psi$ is a text encoder (e.g., Sentence-BERT \cite{reimers-gurevych-2019-sentence}, MiniLM \cite{NEURIPS2020_3f5ee243}) producing embeddings for $q$ and node descriptions $v$.
$g_\phi$ is a fusion function generating node-specific contextual representations $\widetilde{\mathbf{H}}_v$ from node embeddings and the latent variable $\mathbf{H}$.
$\tau$ is a temperature parameter.
Nodes are sampled independently based on $\pi_v$ to form the set $\mathcal{V}$ (typically of size $d$).

\subsection{Edge Optimizer}
\label{sec_edge}

Given the selected nodes $\mathcal{V}$, determined by Section~\ref{sec_node}, the Edge Optimizer determines the set of edges $\mathcal{E}$, including both inter-node connections with strategies ($s_{eg} \in \mathbb{S}_{edge}$) and self-loops with internal strategies ($s_{sl} \in \mathbb{S}_{self}$).
Unlike previous work \cite{gptswarm2024}, which models edges by assigning a real value representing the likelihood of their existence, our approach extends this representation by assigning a probability distribution to each edge. 
This distribution captures not only the probability of edge existence but also the likelihood of selecting a particular strategy type. 
We model the conditional probability distribution $\mathbb{F}_{\theta_e}(\mathcal{E} \mid q, \mathcal{V})$ factorized over potential edges:
\begin{flalign}
    & \nonumber \mathbb{F}_{\theta_e}(\mathcal{E} \mid q, \mathcal{V}) = \\ 
    & \prod_{u, v \in \mathcal{V}, u\ne v} \pi_{eg}( (u, v, s_{eg}) \mid q, u, v, \mathcal{E}_{prev}) \cdot \\
    & \nonumber \prod_{v \in \mathcal{V}} \pi_{sl}( (v, v, s_{sl}) \mid q, v, \mathcal{E}_{prev})
\end{flalign}
where $\pi_{eg}$ is the probability for inter-node edge $(u, v, s_{eg})$ and $\pi_{sl}$ is the probability for self-loop $(v, v, s_{sl})$.
These probabilities depend on contextual node representations $\widetilde{\mathbf{H}}_u$, $\widetilde{\mathbf{H}}_v$ (from Eq.~\ref{eq:node_prob}) and strategy embeddings $\mathbf{H}_{eg}$, $\mathbf{H}_{sl}$:
\begin{flalign}
    & \nonumber \pi_{eg}((u, v, s_{eg}) \mid q, u, v,  \mathcal{E}_{prev}) \propto 
\\ 
    & \begin{cases} exp(\widetilde{\mathbf{H}}_{u,v} \mathbf{H}_{eg}^\top / \tau) & \text{if it's DAG},
    \\ 0 & \text{otherwise}. \end{cases} 
\\
    & \nonumber \pi_{sl}((v, v, s_{sl}) \mid q, v, \mathcal{E}_{prev}) \propto exp( \widetilde{\mathbf{H}}_{v} \mathbf{H}_{sl}^\top / \tau)
\end{flalign}
where $\widetilde{\mathbf{H}}_{u,v} = \text{FFN}(\widetilde{\mathbf{H}}_u || \widetilde{\mathbf{H}}_v)$ denotes the latent representation capturing the implicit relationship between nodes $u$ and $v$ under the context of query $q$. 
The terms $\mathbf{H}_{eg} = \text{FFN}(f_\psi(s_{eg}))$ and $\mathbf{H}_{sl} = \text{FFN}(f_\psi(s_{sl}))$ represent the hidden embeddings for edge strategies and self-loop strategies, respectively.
The crucial DAG constraint ensures that only inter-node edges preserving acyclicity have non-zero probability, effectively pruning cyclic configurations during edge set $\mathcal{E}$ construction.

\begin{table*}[htb]
\centering
\resizebox{\linewidth}{!}{
\begin{tabular}{c|ccc|ccccc}
\hline
\textbf{Method}                          & \textbf{LLM backbone} & MAS  & Routing     & \textbf{MMLU}        & \textbf{GSM8K}       & \textbf{MATH}        & \textbf{HumanEval}   & \textbf{MBPP}        \\ \hline
\multirow{5}{*}{Vanilla} 
& gpt-4o-mini & \usym{2717} & \usym{2717} & 77.81 & 93.17 & 66.09 & 85.71 & 72.20 
\\
& claude-3.5-haiku & \usym{2717} & \usym{2717} & 67.97 & 92.16 & 65.89 & 86.33 & 73.40 
\\
& gemini-1.5-flash & \usym{2717} & \usym{2717} & 80.04 & 92.67 & 74.39 & 82.61 & 73.00 
\\
& llama-3.1-70b & \usym{2717} & \usym{2717} & 79.08 & 92.68 & 60.31 & 80.75 & 68.20 
\\ 
\hline
\multirow{2}{*}{GPTSwarm \cite{gptswarm2024}}       & gpt-4o-mini & \usym{2713} & \usym{2717}      & 82.80       & 94.66       & 68.85       & 86.28       & 75.40      
\\
& gemini-1.5-flash & \usym{2713} & \usym{2717} & 83.22       & 93.98       & 73.35       & 82.36       & 74.80       
\\
\multirow{2}{*}{AgentPrune \cite{zhang2025cut}}     & gpt-4o-mini & \usym{2713} & \usym{2717}      & 83.02       & 94.89       & 68.45       & 86.80       & 75.40       
\\
 & gemini-1.5-flash & \usym{2713} & \usym{2717} & 83.10       & 93.88       & 73.54       & 82.55       & 75.80       
 \\
\multirow{2}{*}{AFlow \cite{zhang2025aflow}}          & gpt-4o-mini & \usym{2713} & \usym{2717}      & 83.10       & 92.30       & 73.35       & 90.06       & 82.20       
\\
& gemini-1.5-flash & \usym{2713} & \usym{2717} & 82.35       & 94.91       & 72.70       & 85.69       & 76.00       
\\ 
\hline
PromptLLM \cite{feng2025graphrouter}                       & LLM Pool & \usym{2717} & \usym{2713}         & 78.43       & 93.92       & 73.03       & 86.33       & 73.60       
\\
RouteLLM \cite{ong2025routellm}                        & LLM Pool & \usym{2717} & \usym{2713}         & 81.04       & 93.42       & 71.29       & 83.85       & 72.60       
\\
FrugalGPT \cite{chen2024frugalgpt}                       & LLM Pool & \usym{2717} & \usym{2713}         & 76.24       & 90.76       & 67.05       & 87.31       & 74.40       
\\
RouterDC \cite{chen2024routerdc}                       & LLM Pool & \usym{2717} & \usym{2713}         & 82.01       & 93.68       & 73.46       & 87.75       & 75.20       
\\ 
\hline
MasRouter \cite{yue2025masrouter}                       & LLM Pool & \usym{2713} & \usym{2713}  & 84.25 & 95.45 & 75.42 & 90.62 & 84.00 
\\
SC-MAS (Ours)   & LLM Pool & \usym{2713} & \usym{2713}   & \textbf{87.60}  & \textbf{96.09}  & \textbf{76.75}  & \textbf{92.37}  & \textbf{87.53}  
\\ 
\hline
\end{tabular}
}
\caption{Performance comparison across multiple categories of baseline methods, including dynamic multi-agent, single-agent routing and multi-agent routing methods. All models draw from the same candidate LLM pool specified in Section~\ref{sec::exp-setup} to ensure a fair comparison.}
\label{tab::perfence-main}
\end{table*}

\subsection{LLM Router}
\label{sec_llm}

Prior studies \cite{chen2024frugalgpt, ong2025routellm, yue2025masrouter} have demonstrated that LLM routing strategies can significantly reduce API costs while maintaining response quality.
Finally, the objective of LLM Router is to assign a suitable LLM $l \in \mathbb{L}$ to each selected agent $v \in \mathcal{V}$, considering the established graph structure $\mathcal{E}$.
Leveraging the semantic edge strategies, we use a Graph Neural Network (GNN) \cite{ZHOU202057} to capture structural context. 
An adjacency matrix $\mathbf{A}$ is constructed where $\mathbf{A}_{u,v} = f_s(f_\psi(s))$ if $(u, v, s) \in \mathcal{E}$ (and $\mathbf{A}_{u,v} = 0$ otherwise), using an edge weighting function $f_s: \mathbb{R}^D \to \mathbb{R}$ on strategy embeddings.
Applying graph convolution layers:
\begin{align}
    \widehat{\mathbf{H}}^{(k)} = \sigma(\widetilde{\mathbf{D}}^{-\frac{1}{2}} \widetilde{\mathbf{A}} \widetilde{\mathbf{D}}^{-\frac{1}{2}} \widehat{\mathbf{H}}^{(k-1)} \mathbf{W}^{(k-1)})
\end{align}
where $\widetilde{\mathbf{A}} = \mathbf{A} + \mathbf{I}$, $\widetilde{\mathbf{D}} = \text{diag}(\sum_{j} \widetilde{\mathbf{A}}_{i, j})$ is the degree matrix, $\sigma$ is an activation function, $\mathbf{W}^{(k)}$ are learnable weights, and initial representations $\widehat{\mathbf{H}}^{(0)}$ are derived from $\{ \widetilde{\mathbf{H}}_v \mid v \in \mathcal{V} \}$.
The final GNN node representations $h_v = \widehat{\mathbf{H}}^{(K)}_v$ encode structural and collaborative context.
The LLM assignment probability for node $v$ is then:
\begin{equation}
\begin{gathered}
    \mathbb{F}_{\theta_l}(\mathcal{L} \mid q, \mathcal{V}, \mathcal{E}) = \prod_{v \in \mathcal{V}} \pi(l \mid q, h_v) \\
\pi(l \mid q, h_v) \propto exp({\text{FFN}(h_v; f_\psi(q))}^\top \mathbf{H}_l / \tau), \\ \tau > 0
\end{gathered}
\end{equation}
where $\mathbf{H}_l = \text{FFN}(f_\psi(l))$ is the embedding for candidate LLM $l$.
This allows assigning LLMs based on both the node's role (via $h_v$) and the query context.

\subsection{Optimization}
\label{sec_opt}

The parameters $\theta$ governing the \textbf{SC-MAS} policy $\mathbb{F}_\theta(\mathcal{G} \mid q)$ are optimized to enhance task performance while considering execution costs. 
Specifically, we aim to maximize the expected log-likelihood of generating the correct answer $a$ for a given query $q$, penalized by the cost associated with the executed graph $\mathcal{G}$.
The optimization objective is formulated as:
\begin{align}
    \min_{\theta} \mathbf{E}_{\underset{\mathcal{G} \sim \mathbb{F}_\theta}{(q, a) \sim \mathcal{D},} }
    [
    -p(a \mid q) + \lambda \cdot C(\mathcal{G}; q)
    ]
\label{eq:opt}
\end{align}
Here, $C(\mathcal{G}; q)$ represents the execution cost, $\lambda \ge 0$ is a hyperparameter balancing performance and cost. 
$p(a \mid q)$ is defined as follows:
\begin{equation}
\begin{gathered}
    p(a \mid q) = \int \mathcal{O}(a \mid \mathcal{G}) \mathbb{F}_\theta(\mathcal{G} \mid q) d\mathcal{G}, \\ \mathcal{G} = (\mathcal{V}, \mathcal{E}, \mathcal{L})
\end{gathered}
\end{equation}
where $\mathcal{O}(a \mid \mathcal{G})$ denotes the likelihood of obtaining answer $a$ by executing $\mathcal{G}$. $\mathbb{F}_\theta(\mathcal{G} \mid q)$ was computed in the previous sections.
To optimize this objective (Eq.~\ref{eq:opt}), we employ policy gradient \cite{NIPS1999_6449f44a, williams1992simple}, following common practices in multi-agent structure optimization \cite{gptswarm2024, zhang2025aflow, yue2025masrouter}.
A summary of notations is available in Appendix~\ref{app::notation}, with the detailed algorithm presented in Appendix~\ref{app::algorithm-workflow}.

\section{Experiments}
\label{sec::experiments}

\begin{table*}[htbp]
\resizebox{\linewidth}{!}{
\begin{tabular}{c|cc|cc|cc|cc|cc}
\hline
\multirow{2}{*}{\textbf{Dataset}} & \multicolumn{2}{c|}{\textbf{MMLU}} & \multicolumn{2}{c|}{\textbf{GSM8K}} & \multicolumn{2}{c|}{\textbf{MATH}} & \multicolumn{2}{c|}{\textbf{HumanEval}} & \multicolumn{2}{c}{\textbf{MBPP}} \\ \cline{2-11} 
                         & Acc       & Cost (\$)     & Acc        & Cost (\$)     & Acc       & Cost (\$)     & Acc          & Cost (\$)       & Acc       & Cost (\$)    \\ \hline
vanilla SC-MAS               & 87.60     & 1.21          & 96.09      & 1.33          & 76.75     & 3.18          & 92.37        & 0.15           & 87.53     & 0.91         \\ \hline
w/o $\mathbb{F}_{\theta_v}$                      & 85.46     & 1.29          & 95.30      & 1.47          & 74.37     & 3.20          & 87.02        & 0.16           & 83.63     & 0.90         \\
w/o $\mathbb{F}_{\theta_e}$                      & 86.17     & 1.71          & 94.92      & 1.82          & 73.03     & 3.71          & 88.54        & 0.18           & 84.50     & 1.17         \\
w/o $\mathbb{F}_{\theta_l}$                      & 85.96     & 1.44          & 94.84      & 1.44          & 74.95     & 3.85          & 85.50        & 0.16           & 85.09     & 1.29         \\ \hline
\end{tabular}
}
\caption{Ablation study of SC-MAS, w/o stands for replacing it with randomized selection.}
\label{tab::app-alblation}
\end{table*}

\subsection{Experimental Setup}
\label{sec::exp-setup}

\paragraph{Datasets}
To comprehensively evaluate \textbf{SC-MAS} across a diverse set of capabilities, we utilize five widely adopted benchmark datasets: \textbf{MMLU} \cite{hendrycks2021measuring}, \textbf{GSM8K} \cite{cobbe2021trainingverifierssolvemath}, \textbf{MATH} \cite{hendrycks2021measuringmath}, \textbf{HumanEval} \cite{chen2021evaluatinglargelanguagemodels}, and \textbf{MBPP} \cite{austin2021programsynthesislargelanguage}. These datasets collectively encompass a broad range of task domains, including general knowledge, mathematical problem solving, and program synthesis.
Following standard practice, we partition each dataset into validation and test subsets using a 1:4 ratio.
Specifically for the MATH dataset, we adopt the protocol established by prior work \cite{yue2025masrouter} and perform stratified sampling to construct a subset of 519 problems spanning multiple difficulty levels.

\paragraph{Baselines}
We compare \textbf{SC-MAS} against an extensive set of baselines grouped into three categories:
(1) dynamic multi-agent systems, such as GPTSwarm \cite{gptswarm2024}, AgentPrune \cite{zhang2025cut} and AFlow \cite{zhang2025aflow}; 
(2) single-agent LLM routing methods, including PromptLLM \cite{feng2025graphrouter}, RouteLLM \cite{ong2025routellm}, FrugalGPT \cite{chen2024frugalgpt} and RouterDC \cite{chen2024routerdc}. 
(3) multi-agent system routing method, represented by MasRouter \cite{yue2025masrouter}.

\paragraph{LLM Backbones}
For consistency and fairness in comparison, we adopt the same set of LLM backbones as used in previous work \cite{yue2025masrouter}, comprising \texttt{gpt-4o-mini-0718} \cite{openaiGPT4oMini}, \texttt{claude-3.5-haiku} \cite{claude35haiku}, \texttt{gemini-1.5-flash} \cite{geminiteam2024gemini15unlockingmultimodal}, and \texttt{llama-3.1-70b} \cite{grattafiori2024llama3herdmodels}.
For ablation studies, we additionally include \texttt{Deepseek-v3} \cite{deepseekai2025deepseekv3technicalreport}. 
The temperature parameters remain at default settings.

\paragraph{Implementation Details}
We adopt all LLMs listed in the LLM Backbones as the candidate model pool for routing methods.
To model inter-agent collaboration, we include a diverse edge strategy repository consisting of Chain \cite{qian2025scaling}, Debate \cite{chan2023chateval}, and Criticism \cite{huang-chang-2023-towards}.
For self-loop strategies, we employ CoT \cite{cot2022} and Reflection \cite{shinn2023reflexion}.
The role pool comprises 26 distinct agent roles, following the configuration proposed by MasRouter \cite{yue2025masrouter}, and spans a wide range of functional expertise across coding, mathematical reasoning, and commonsense domains.
We set the learning rate $\alpha = 0.01$ and the temperature parameter $\tau = 1$. 
We run the experiment three times on each dataset and report the average as the final result.
To examine the influence of cost sensitivity in routing decisions, we vary the cost penalty coefficient $\lambda$ within the set $\{0, 5, 10, 15, 20\}$. The report results use $\lambda$ as 5.

\subsection{Performance \& Cost Analysis}
In this section, we compare \textbf{SC-MAS} against eight competitive baselines across five representative benchmarks. Our evaluation demonstrates that \textbf{SC-MAS} is:

\textbf{High Performance}.
The experimental results, as detailed in Table~\ref{tab::perfence-main}.
Our method achieves SOTA results across all five datasets, including a improvement of 3.35\% on the MMLU and 3.53\% on MBPP compared to MasRouter.
These findings underscore the benefits of incorporating heterogeneous collaboration strategies within MAS, enabling more adaptive and efficient problem solving.

\begin{figure}[htp]
    \centering
    \includegraphics[width=1.\linewidth]{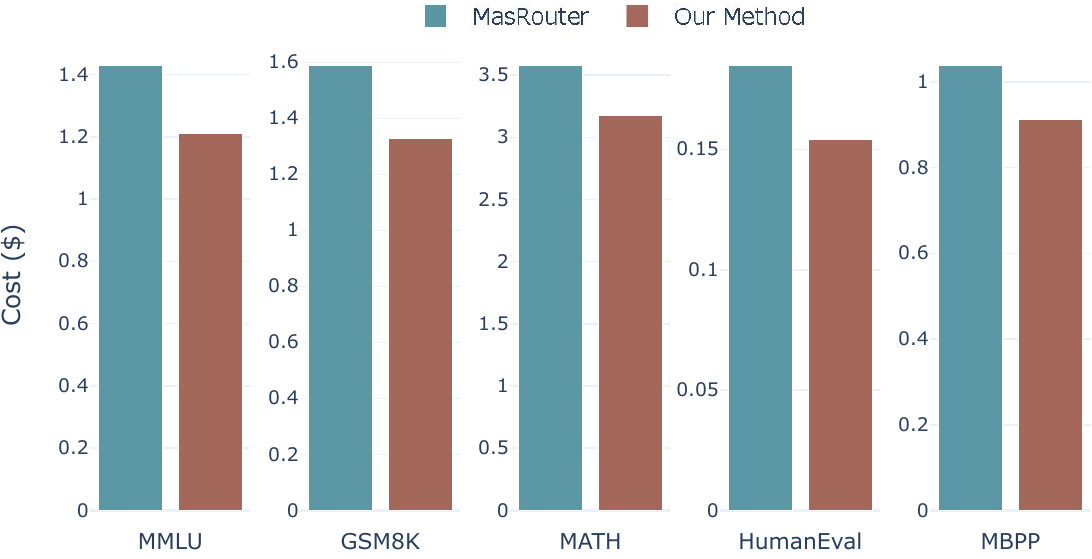}
    \caption{Comparison of inference costs between our work and MasRouter \cite{yue2025masrouter} across various datasets.}
\label{fig::infer-cost}
\end{figure}

\textbf{Inference Token Efficiency}.
As illustrated in Figure~\ref{fig::infer-cost}, \textbf{SC-MAS} demonstrates enhanced cost-effectiveness.
Compared to MasRouter, our method achieves a 3.53\% performance gain while concurrently reducing inference overhead by 12.13\% on the MBPP dataset.
Similarly, on the HumanEval dataset, it achieves a 1.75\% performance gain accompanied by a 16.76\% reduction in inference cost.
These outcomes underscore the capacity of \textbf{SC-MAS} to effectively balance performance with cost.

\begin{figure}[htp]
    \centering
    \includegraphics[width=1.\linewidth]{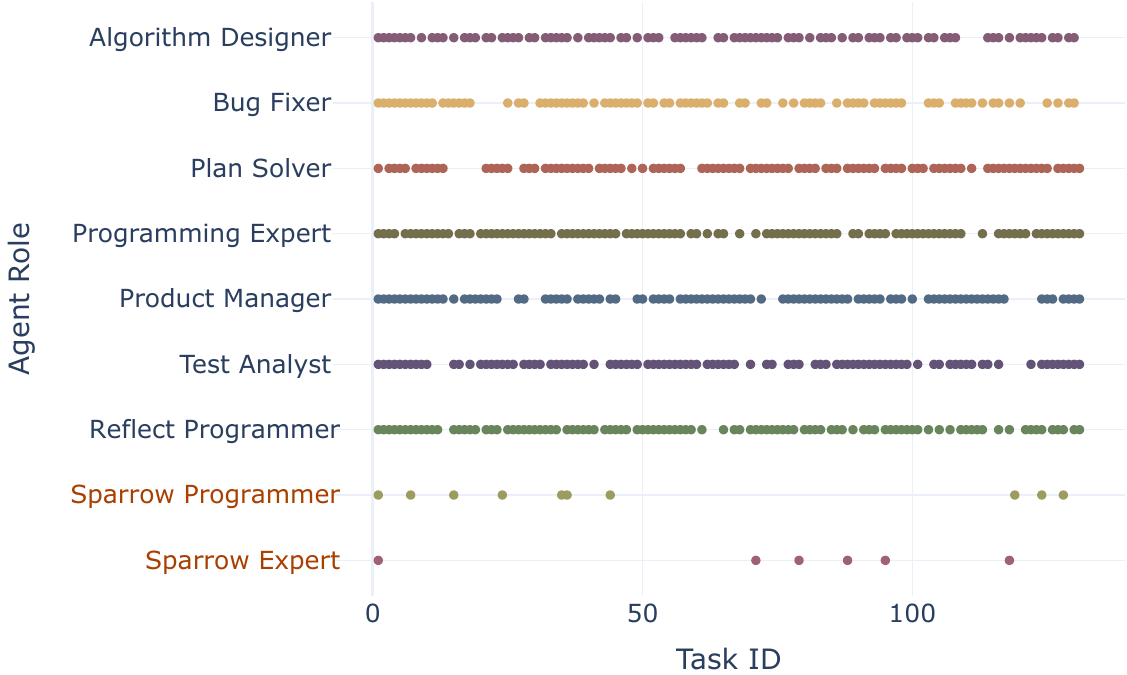}
		\caption{Distribution of roles chosen for reasoning on the HumanEval dataset, with the two roles beginning with \textit{Sparrow} being the agents coded as error outputs.}
\label{fig::adversarial}
\end{figure}

\subsection{Ablation Study}

We perform an ablation study to evaluate the contributions of three critical components of \textbf{SC-MAS}. 
The results, summarized in Table~\ref{tab::app-alblation}, highlight the importance of each module to overall system effectiveness and efficiency.
Removing $\mathbb{F}_{\theta_e}$ resulted in a performance drop of 1.17\% to 3.83\% and an accompanying cost increase of 16.67\% to 41.32\%, which underscores the importance of capturing latent relationships among agents for efficient collaboration.
Meanwhile, removing $\mathbb{F}_{\theta_l}$ reduced performance by 1.25\% to 6.87\% and led to a cost increase of up to 41.76\%, indicating that combining diverse LLMs contributes meaningfully to both performance enhancement and cost efficiency.

\subsection{Analysis of Unhelpful Agent Filtering}

In realistic multi-agent settings, some candidate agents may generate irrelevant or low-quality responses due to role mismatch or limited capability. A practical multi-agent system should be able to reduce the influence of such unhelpful collaborators.
To examine this property, we augment the agent pool with two additional roles that deliberately produce random and task-irrelevant outputs. These agents are not designed to perform targeted attacks, but instead act as sources of noisy information.
As shown in Figure~\ref{fig::adversarial}, \textbf{SC-MAS} assigns significantly lower selection probabilities to these unhelpful agents (7.63\% and 4.58\%) compared to task-relevant roles. This suggests that \textbf{SC-MAS} can implicitly filter low-quality collaborators during agent selection.

\begin{figure}[htbp]
    \centering
    \includegraphics[width=1.\linewidth]{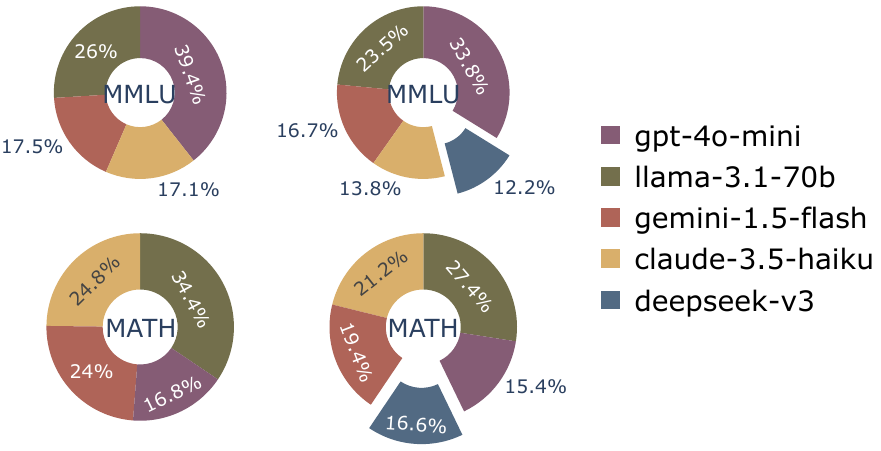}
		\caption{The selected LLM distribution of SC-MAS on MMLU and MATH.}
		\label{fig::inductive}
\end{figure}

\begin{figure}[htbp]
    \centering
    \includegraphics[width=1.\linewidth]{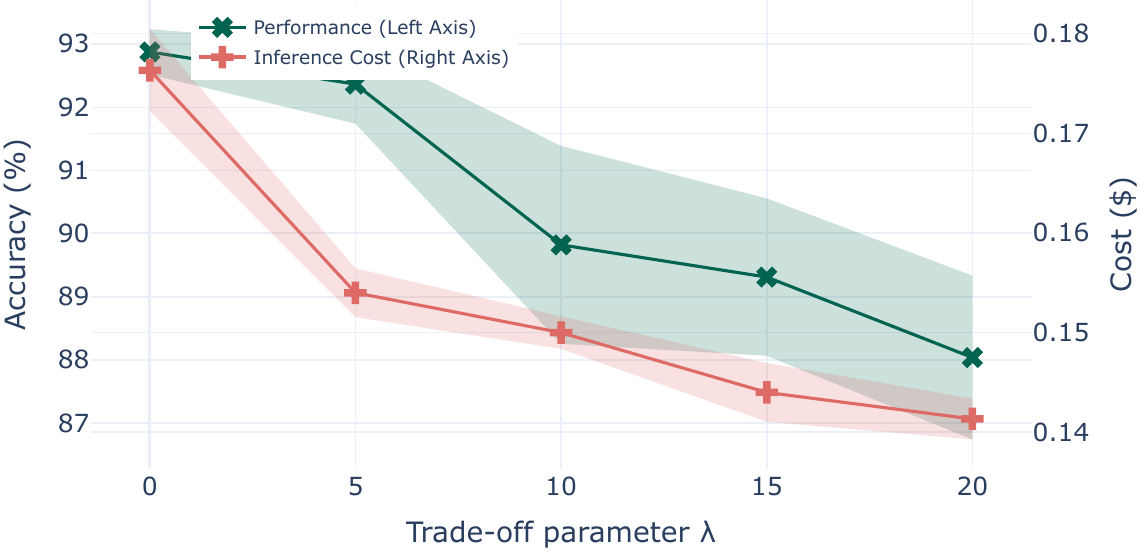}
		\caption{Sensitivity analysis of trade-off parameter $\lambda$ on HumanEval.}
		\label{fig::sensitivity}
\end{figure}

\subsection{Inductive Ability Analysis}

In this section, we demonstrate that \textbf{SC-MAS} can generalize to previously unseen LLMs without retraining. 
As illustrated in Figure~\ref{fig::inductive}, the newly introduced model \texttt{Deepseek-v3} is selected 12.2\% and 16.6\% of the time on MMLU and MATH, respectively.
After incorporating this stronger LLM, \textbf{SC-MAS} achieves increased accuracy on MMLU from 87.60\% to 88.14\%, and on MATH from 76.75\% to 77.84\%.

\subsection{Sensitivity Analysis}
\label{sec::sensitivity}

In this section, we perform a sensitivity analysis of the $\lambda$ parameter introduced in Eq.~\ref{eq:opt} to balance performance and cost.
As $\lambda$ increases from 0 to 20, the results in Figure~\ref{fig::sensitivity} illustrate the corresponding trade-offs.
When $\lambda$ is 0, the optimization objective considers only performance, resulting in higher overhead.
As $\lambda$ grows, \textbf{SC-MAS} begins to favor more cost-effective solutions, leading to a reduction in overhead by 19.89\%, at the expense of a 4.84\% performance drop.
Thus, adjusting $\lambda$ according to available resources and task requirements effectively balances performance and cost.

\subsection{Case Study}

We provide case studies and visualizations of the MAS generated by SC-MAS across five benchmarks in Appendix~\ref{app::case-study}.

\section{Conclusion}
\label{sec::conclusion}

In this paper, we study query-conditioned multi-agent system design from a construction perspective and introduce a social capital-driven framework for adaptive and cost-efficient MAS.
We propose \textbf{SC-MAS}, a high-performance and cost-effective framework for heterogeneous MAS construction.
Through its \textit{Node Selector}, \textit{Edge Optimizer}, and \textit{LLM Assigner} modules, \textbf{SC-MAS} progressively constructs an executable MAS that balances task performance and computational cost.
Our results demonstrate that explicitly modeling edge-level collaboration and resource allocation provides an effective and principled solution for adaptive multi-agent system construction, offering new insights for scalable and flexible LLM-based MAS design.

\section{Limitation}
\label{sec::limitation}

\paragraph{Structural Constraints on MAS Topology}
To ensure orderly and terminable execution, \textbf{SC-MAS} imposes a directed acyclic graph (DAG) structure on the multi-agent system, consistent with prior work.
While effective in practice, this constraint inherently limits the expressiveness of the system, precluding cyclical communication patterns and feedback loops that may be beneficial for tasks requiring iterative refinement or mutual verification.
This could be a direction for follow-up research.

\paragraph{Limited Interpretability of Agent and Strategy Selection}
Additionally, although \textbf{SC-MAS} automatically generates MAS configurations that demonstrate good performance, the rationale behind selecting specific agents, collaboration strategies, and large language models remains non-trivial to interpret. 
Strengthening the interpretability of MAS generation would enhance user understanding and trust, representing a valuable direction for future research.


\bibliography{custom}

@inproceedings{cot2022,
author = {Wei, Jason and Wang, Xuezhi and Schuurmans, Dale and Bosma, Maarten and Ichter, et al},
title = {Chain-of-thought prompting elicits reasoning in large language models},
year = {2022},
isbn = {9781713871088},
publisher = {Curran Associates Inc.},
address = {Red Hook, NY, USA},
booktitle = {Proceedings of the 36th International Conference on Neural Information Processing Systems},
articleno = {1800},
numpages = {14},
location = {New Orleans, LA, USA},
series = {NIPS '22}
}

@inproceedings{
du2024improving,
title={Improving Factuality and Reasoning in Language Models through Multiagent Debate},
author={Yilun Du and Shuang Li and Antonio Torralba and Joshua B. Tenenbaum and Igor Mordatch},
booktitle={Forty-first International Conference on Machine Learning},
year={2024},
url={https://openreview.net/forum?id=zj7YuTE4t8}
}

@inproceedings{liang-etal-2024-encouraging,
    title = "Encouraging Divergent Thinking in Large Language Models through Multi-Agent Debate",
    author = "Liang, Tian  and
      He, Zhiwei  and
      Jiao, Wenxiang  and
      Wang, et al",
    editor = "Al-Onaizan, Yaser  and
      Bansal, Mohit  and
      Chen, Yun-Nung",
    booktitle = "Proceedings of the 2024 Conference on Empirical Methods in Natural Language Processing",
    month = nov,
    year = "2024",
    address = "Miami, Florida, USA",
    publisher = "Association for Computational Linguistics",
    url = "https://aclanthology.org/2024.emnlp-main.992/",
    doi = "10.18653/v1/2024.emnlp-main.992",
    pages = "17889--17904"
}

@article{shinn2023reflexion,
  title={Reflexion: Language agents with verbal reinforcement learning},
  author={Shinn, Noah and Cassano, Federico and Gopinath, Ashwin and Narasimhan, Karthik and Yao, Shunyu},
  journal={Advances in Neural Information Processing Systems},
  volume={36},
  pages={8634--8652},
  year={2023}
}

@inproceedings{gptswarm2024,
author = {Zhuge, Mingchen and Wang, Wenyi and Kirsch, Louis and Faccio, et al},
title = {GPTSwarm: language agents as optimizable graphs},
year = {2024},
publisher = {JMLR.org},
booktitle = {Proceedings of the 41st International Conference on Machine Learning},
articleno = {2597},
numpages = {25},
location = {Vienna, Austria},
series = {ICML'24}
}

@inproceedings{
qian2025scaling,
title={Scaling Large Language Model-based Multi-Agent Collaboration},
author={Chen Qian and Zihao Xie and YiFei Wang and Wei Liu and Kunlun Zhu and et al},
booktitle={The Thirteenth International Conference on Learning Representations},
year={2025},
url={https://openreview.net/forum?id=K3n5jPkrU6}
}

@misc{wang2025agentdropout,
      title={AgentDropout: Dynamic Agent Elimination for Token-Efficient and High-Performance LLM-Based Multi-Agent Collaboration}, 
      author={Zhexuan Wang and Yutong Wang and Xuebo Liu and Liang Ding and Miao Zhang and Jie Liu and Min Zhang},
      year={2025},
      eprint={2503.18891},
      archivePrefix={arXiv},
      primaryClass={cs.CL},
      url={https://arxiv.org/abs/2503.18891}, 
}

@inproceedings{yue2025masrouter,
    title = "{M}as{R}outer: Learning to Route {LLM}s for Multi-Agent Systems",
    author = "Yue, Yanwei  and
      Zhang, Guibin  and
      Liu, Boyang  and
      Wan, Guancheng  and
      Wang, Kun  and
      Cheng, Dawei  and
      Qi, Yiyan",
    editor = "Che, Wanxiang  and
      Nabende, Joyce  and
      Shutova, Ekaterina  and
      Pilehvar, Mohammad Taher",
    booktitle = "Proceedings of the 63rd Annual Meeting of the Association for Computational Linguistics (Volume 1: Long Papers)",
    month = jul,
    year = "2025",
    address = "Vienna, Austria",
    publisher = "Association for Computational Linguistics",
    url = "https://aclanthology.org/2025.acl-long.757/",
    doi = "10.18653/v1/2025.acl-long.757",
    pages = "15549--15572",
    ISBN = "979-8-89176-251-0"
}

@article{coleman1988social,
  title={Social capital in the creation of human capital},
  author={Coleman, James S},
  journal={American journal of sociology},
  volume={94},
  pages={S95--S120},
  year={1988},
  publisher={University of Chicago Press}
}

@article{putnam2001social,
  title={Social capital: Measurement and consequences},
  author={Putnam, Robert and others},
  journal={Canadian journal of policy research},
  volume={2},
  number={1},
  pages={41--51},
  year={2001},
  publisher={Saint-Laurent, Qu{\'e}bec}
}

@article{chen2024frugalgpt,
  title={FrugalGPT: How to Use Large Language Models While Reducing Cost and Improving Performance},
  author={Chen, Lingjiao and Zaharia, Matei and Zou, James},
  journal={Transactions on Machine Learning Research},
  year={2024}
}

@inproceedings{
ong2025routellm,
title={Route{LLM}: Learning to Route {LLM}s from Preference Data},
author={Isaac Ong and Amjad Almahairi and Vincent Wu and Wei-Lin Chiang and et al},
booktitle={The Thirteenth International Conference on Learning Representations},
year={2025},
url={https://openreview.net/forum?id=8sSqNntaMr}
}

@article{ZHOU202057,
title = {Graph neural networks: A review of methods and applications},
journal = {AI Open},
volume = {1},
pages = {57-81},
year = {2020},
issn = {2666-6510},
doi = {https://doi.org/10.1016/j.aiopen.2021.01.001},
url = {https://www.sciencedirect.com/science/article/pii/S2666651021000012},
author = {Jie Zhou and Ganqu Cui and Shengding Hu and Zhengyan Zhang and Cheng Yang and et al}
}

@inproceedings{
zhang2025aflow,
title={{AF}low: Automating Agentic Workflow Generation},
author={Jiayi Zhang and Jinyu Xiang and Zhaoyang Yu and Fengwei Teng and et al},
booktitle={The Thirteenth International Conference on Learning Representations},
year={2025},
url={https://openreview.net/forum?id=z5uVAKwmjf}
}

@inproceedings{NIPS1999_6449f44a,
 author = {Konda, Vijay and Tsitsiklis, John},
 booktitle = {Advances in Neural Information Processing Systems},
 editor = {S. Solla and T. Leen and K. M\"{u}ller},
 pages = {},
 publisher = {MIT Press},
 title = {Actor-Critic Algorithms},
 url = {https://proceedings.neurips.cc/paper_files/paper/1999/file/6449f44a102fde848669bdd9eb6b76fa-Paper.pdf},
 volume = {12},
 year = {1999}
}

@inproceedings{guo2024large,
author = {Guo, Taicheng and Chen, Xiuying and Wang, Yaqi and Chang, et al},
title = {Large language model based multi-agents: a survey of progress and challenges},
year = {2024},
isbn = {978-1-956792-04-1},
url = {https://doi.org/10.24963/ijcai.2024/890},
doi = {10.24963/ijcai.2024/890},
articleno = {890},
numpages = {10},
location = {Jeju, Korea},
series = {IJCAI '24}
}

@inproceedings{
hong2023metagpt,
title={Meta{GPT}: Meta Programming for A Multi-Agent Collaborative Framework},
author={Sirui Hong and Mingchen Zhuge and Jonathan Chen and Xiawu Zheng and et al},
booktitle={The Twelfth International Conference on Learning Representations},
year={2024},
url={https://openreview.net/forum?id=VtmBAGCN7o}
}

@inproceedings{
chan2023chateval,
title={ChatEval: Towards Better {LLM}-based Evaluators through Multi-Agent Debate},
author={Chi-Min Chan and Weize Chen and Yusheng Su and Jianxuan Yu and et al},
booktitle={The Twelfth International Conference on Learning Representations},
year={2024},
url={https://openreview.net/forum?id=FQepisCUWu}
}

@inproceedings{li2023camel,
author = {Li, Guohao and Al Kader Hammoud, Hasan Abed and Itani, Hani and Khizbullin, Dmitrii and Ghanem, Bernard},
title = {CAMEL: communicative agents for "mind" exploration of large language model society},
year = {2023},
publisher = {Curran Associates Inc.},
address = {Red Hook, NY, USA},
booktitle = {Proceedings of the 37th International Conference on Neural Information Processing Systems},
articleno = {2264},
numpages = {18},
location = {New Orleans, LA, USA},
series = {NIPS '23},
url={https://ghli.org/publication/neurips2023camel/}
}

@inproceedings{park2023generative,
  title={Generative agents: Interactive simulacra of human behavior},
  author={Park, Joon Sung and O'Brien, Joseph and Cai, Carrie Jun and Morris, et al},
  booktitle={Proceedings of the 36th annual acm symposium on user interface software and technology},
  pages={1--22},
  year={2023},
  url={https://dl.acm.org/doi/10.1145/3586183.3606763}
}

@article{gao2023s,
  title={S3: Social-network Simulation System with Large Language Model-Empowered Agents},
  author={Chen Gao and Xiaochong Lan and Zhi-jie Lu and Jinzhu Mao and et al},
  journal={ArXiv},
  year={2023},
  volume={abs/2307.14984},
  url={https://api.semanticscholar.org/CorpusID:260202947}
}

@misc{
akata2023playing,
title={Playing repeated games with Large Language Models},
author={Elif Akata and Lion Schulz and Julian Coda-Forno and Seong Joon Oh and Matthias Bethge and Eric Schulz},
year={2024},
url={https://openreview.net/forum?id=CSpWgKo0ID}
}

@article{wang2023voyager,
  title={Voyager: An Open-Ended Embodied Agent with Large Language Models},
  author={Guanzhi Wang and Yuqi Xie and Yunfan Jiang and et al},
  journal={Trans. Mach. Learn. Res.},
  year={2023},
  volume={2024},
  url={https://api.semanticscholar.org/CorpusID:258887849}
}

@inproceedings{
tan2024towards,
title={Towards General Computer Control: A Multimodal Agent for Red Dead Redemption {II} as a Case Study},
author={Weihao Tan and Ziluo Ding and Wentao Zhang and Boyu Li and et al},
booktitle={ICLR 2024 Workshop on Large Language Model (LLM) Agents},
year={2024},
url={https://openreview.net/forum?id=pmcFzuUxsP}
}

@inproceedings{qian-etal-2024-chatdev,title = "{C}hat{D}ev: Communicative Agents for Software Development",author = "Qian, Chen and Liu, Wei and Liu, Hongzhang and Chen, et al",editor = "Ku, Lun-Wei and Martins, Andre and Srikumar, Vivek",booktitle = ACL:2024:long,month = aug,year = "2024",address = "Bangkok, Thailand",publisher = acl,url = anth # {2024.acl-long.810/},doi = "10.18653/v1/2024.acl-long.810",pages = "15174--15186"
}

@misc{matarazzo2025surveylargelanguagemodels,
      title={A Survey on Large Language Models with some Insights on their Capabilities and Limitations}, 
      author={Andrea Matarazzo and Riccardo Torlone},
      year={2025},
      eprint={2501.04040},
      archivePrefix={arXiv},
      primaryClass={cs.CL},
      url={https://arxiv.org/abs/2501.04040}, 
}

@misc{abdin2024phi3technicalreporthighly,
      title={Phi-3 Technical Report: A Highly Capable Language Model Locally on Your Phone}, 
      author={Marah Abdin and Jyoti Aneja and Hany Awadalla and et al},
      year={2024},
      eprint={2404.14219},
      archivePrefix={arXiv},
      primaryClass={cs.CL},
      url={https://arxiv.org/abs/2404.14219}, 
}

@inproceedings{lepagnol-etal-2024-small,
    title = "Small Language Models Are Good Too: An Empirical Study of Zero-Shot Classification",
    author = "Lepagnol, Pierre  and
      Gerald, Thomas  and
      Ghannay, et al",
    editor = "Calzolari, Nicoletta  and
      Kan, Min-Yen  and
      Hoste, Veronique  and
      Lenci, Alessandro  and
      Sakti, Sakriani  and
      Xue, Nianwen",
    booktitle = "Proceedings of the 2024 Joint International Conference on Computational Linguistics, Language Resources and Evaluation (LREC-COLING 2024)",
    month = may,
    year = "2024",
    address = "Torino, Italia",
    publisher = "ELRA and ICCL",
    url = "https://aclanthology.org/2024.lrec-main.1299/",
    pages = "14923--14936",
}

@inproceedings{shen-etal-2024-small,
    title = "Small {LLM}s Are Weak Tool Learners: A Multi-{LLM} Agent",
    author = "Shen, Weizhou  and
      Li, Chenliang  and
      et al",
    editor = "Al-Onaizan, Yaser  and
      Bansal, Mohit  and
      Chen, Yun-Nung",
    booktitle = "Proceedings of the 2024 Conference on Empirical Methods in Natural Language Processing",
    month = nov,
    year = "2024",
    address = "Miami, Florida, USA",
    publisher = "Association for Computational Linguistics",
    url = "https://aclanthology.org/2024.emnlp-main.929/",
    doi = "10.18653/v1/2024.emnlp-main.929",
    pages = "16658--16680",
}

@inproceedings{
hu2024routerbench,
title={RouterBench: A Benchmark for Multi-{LLM} Routing System},
author={Qitian Jason Hu and Jacob Bieker and Xiuyu Li and et al},
booktitle={Agentic Markets Workshop at ICML 2024},
year={2024},
url={https://openreview.net/forum?id=IVXmV8Uxwh}
}

@inproceedings{devlin-etal-2019-bert,
    title = "{BERT}: Pre-training of Deep Bidirectional Transformers for Language Understanding",
    author = "Devlin, Jacob  and
      Chang, Ming-Wei  and
      et al",
    editor = "Burstein, Jill  and
      Doran, Christy  and
      Solorio, Thamar",
    booktitle = "Proceedings of the 2019 Conference of the North {A}merican Chapter of the Association for Computational Linguistics: Human Language Technologies, Volume 1 (Long and Short Papers)",
    month = jun,
    year = "2019",
    address = "Minneapolis, Minnesota",
    publisher = "Association for Computational Linguistics",
    url = "https://aclanthology.org/N19-1423/",
    doi = "10.18653/v1/N19-1423",
    pages = "4171--4186",
}

@inproceedings{
feng2025graphrouter,
title={GraphRouter: A Graph-based Router for {LLM} Selections},
author={Tao Feng and Yanzhen Shen and Jiaxuan You},
booktitle={The Thirteenth International Conference on Learning Representations},
year={2025},
url={https://openreview.net/forum?id=eU39PDsZtT}
}

@misc{mohammadshahi2024routoolearningroutelarge,
      title={Routoo: Learning to Route to Large Language Models Effectively}, 
      author={Alireza Mohammadshahi and Arshad Rafiq Shaikh and Majid Yazdani},
      year={2024},
      eprint={2401.13979},
      archivePrefix={arXiv},
      primaryClass={cs.CL},
      url={https://arxiv.org/abs/2401.13979}, 
}

@misc{dai2024costeffectiveonlinemultillmselection,
      title={Cost-Effective Online Multi-LLM Selection with Versatile Reward Models}, 
      author={Xiangxiang Dai and Jin Li and et al},
      year={2024},
      eprint={2405.16587},
      archivePrefix={arXiv},
      primaryClass={cs.LG},
      url={https://arxiv.org/abs/2405.16587}, 
}

@article{
li2024more,
title={More Agents Is All You Need},
author={junyou li and Qin Zhang and Yangbin Yu and QIANG FU and Deheng Ye},
journal={Transactions on Machine Learning Research},
issn={2835-8856},
year={2024},
url={https://openreview.net/forum?id=bgzUSZ8aeg},
note={}
}

@inproceedings{
chen2024are,
title={Are More {LLM} Calls All You Need? Towards the Scaling Properties of Compound {AI} Systems},
author={Lingjiao Chen and Jared Quincy Davis and et al},
booktitle={The Thirty-eighth Annual Conference on Neural Information Processing Systems},
year={2024},
url={https://openreview.net/forum?id=m5106RRLgx}
}

@inproceedings{
wang2025mixtureofagents,
title={Mixture-of-Agents Enhances Large Language Model Capabilities},
author={Junlin Wang and Jue WANG and Ben Athiwaratkun and Ce Zhang and James Zou},
booktitle={The Thirteenth International Conference on Learning Representations},
year={2025},
url={https://openreview.net/forum?id=h0ZfDIrj7T}
}

@inproceedings{
    zhang2025cut,
    title={Cut the Crap: An Economical Communication Pipeline for {LLM}-based Multi-Agent Systems},
    author={Guibin Zhang and Yanwei Yue and et al},
    booktitle={The Thirteenth International Conference on Learning Representations},
    year={2025},
    url={https://openreview.net/forum?id=LkzuPorQ5L}
}

@inproceedings{
hu2025automated,
title={Automated Design of Agentic Systems},
author={Shengran Hu and Cong Lu and Jeff Clune},
booktitle={The Thirteenth International Conference on Learning Representations},
year={2025},
url={https://openreview.net/forum?id=t9U3LW7JVX}
}

@misc{zhang2025evoflowevolvingdiverseagentic,
      title={EvoFlow: Evolving Diverse Agentic Workflows On The Fly}, 
      author={Guibin Zhang and Kaijie Chen and Guancheng Wan and Heng Chang and Hong Cheng and Kun Wang and Shuyue Hu and Lei Bai},
      year={2025},
      eprint={2502.07373},
      archivePrefix={arXiv},
      primaryClass={cs.LG},
      url={https://arxiv.org/abs/2502.07373}, 
}

@inproceedings{
shang2025agentsquare,
title={AgentSquare: Automatic {LLM} Agent Search in Modular Design Space},
author={Yu Shang and Yu Li and Keyu Zhao and Likai Ma and Jiahe Liu and Fengli Xu and Yong Li},
booktitle={The Thirteenth International Conference on Learning Representations},
year={2025},
url={https://openreview.net/forum?id=mPdmDYIQ7f}
}

@misc{
liu2024dynamic,
title={Dynamic {LLM}-Agent Network: An {LLM}-agent Collaboration Framework with Agent Team Optimization},
author={Zijun Liu and Yanzhe Zhang and Peng Li and Yang Liu and Diyi Yang},
year={2024},
url={https://openreview.net/forum?id=i43XCU54Br}
}

@inproceedings{
ding2024hybrid,
title={Hybrid {LLM}: Cost-Efficient and Quality-Aware Query Routing},
author={Dujian Ding and Ankur Mallick and Chi Wang and Robert Sim and Subhabrata Mukherjee and Victor R{\"u}hle and Laks V. S. Lakshmanan and Ahmed Hassan Awadallah},
booktitle={The Twelfth International Conference on Learning Representations},
year={2024},
url={https://openreview.net/forum?id=02f3mUtqnM}
}

@inproceedings{
chen2024routerdc,
title={Router{DC}: Query-Based Router by Dual Contrastive Learning for Assembling Large Language Models},
author={Shuhao Chen and Weisen Jiang and Baijiong Lin and James Kwok and Yu Zhang},
booktitle={The Thirty-eighth Annual Conference on Neural Information Processing Systems},
year={2024},
url={https://openreview.net/forum?id=7RQvjayHrM}
}

@misc{
dai2025costeffective,
title={Cost-Effective Online Multi-{LLM} Selection with Versatile Reward Models},
author={Xiangxiang Dai and Jin Li and Xutong Liu and Anqi Yu and John C.S. Lui},
year={2025},
url={https://openreview.net/forum?id=JLDAWbzTUg}
}

@inproceedings{
hendrycks2021measuring,
title={Measuring Massive Multitask Language Understanding},
author={Dan Hendrycks and Collin Burns and Steven Basart and Andy Zou and Mantas Mazeika and Dawn Song and Jacob Steinhardt},
booktitle={International Conference on Learning Representations},
year={2021},
url={https://openreview.net/forum?id=d7KBjmI3GmQ}
}

@misc{cobbe2021trainingverifierssolvemath,
      title={Training Verifiers to Solve Math Word Problems}, 
      author={Karl Cobbe and Vineet Kosaraju and Mohammad Bavarian and Mark Chen and Heewoo Jun and Lukasz Kaiser and Matthias Plappert and Jerry Tworek and Jacob Hilton and Reiichiro Nakano and Christopher Hesse and John Schulman},
      year={2021},
      eprint={2110.14168},
      archivePrefix={arXiv},
      primaryClass={cs.LG},
      url={https://arxiv.org/abs/2110.14168}, 
}

@inproceedings{
hendrycks2021measuringmath,
title={Measuring Mathematical Problem Solving With the {MATH} Dataset},
author={Dan Hendrycks and Collin Burns and Saurav Kadavath and Akul Arora and Steven Basart and Eric Tang and Dawn Song and Jacob Steinhardt},
booktitle={Thirty-fifth Conference on Neural Information Processing Systems Datasets and Benchmarks Track (Round 2)},
year={2021},
url={https://openreview.net/forum?id=7Bywt2mQsCe}
}

@misc{chen2021evaluatinglargelanguagemodels,
      title={Evaluating Large Language Models Trained on Code}, 
      author={Mark Chen and Jerry Tworek and Heewoo Jun and Qiming Yuan and Henrique Ponde de Oliveira Pinto and et al},
      year={2021},
      eprint={2107.03374},
      archivePrefix={arXiv},
      primaryClass={cs.LG},
      url={https://arxiv.org/abs/2107.03374}, 
}

@misc{austin2021programsynthesislargelanguage,
      title={Program Synthesis with Large Language Models}, 
      author={Jacob Austin and Augustus Odena and Maxwell Nye and Maarten Bosma and Henryk Michalewski and David Dohan and Ellen Jiang and Carrie Cai and Michael Terry and Quoc Le and Charles Sutton},
      year={2021},
      eprint={2108.07732},
      archivePrefix={arXiv},
      primaryClass={cs.PL},
      url={https://arxiv.org/abs/2108.07732}, 
}

@misc{openaiGPT4oMini,
	author = {OpenAI},
	title = {{G}{P}{T}-4o mini: advancing cost-efficient intelligence},
	howpublished = {\url{https://openai.com/index/gpt-4o-mini-advancing-cost-efficient-intelligence/}},
	year = {2024},
}

@misc{claude35haiku,
    author = {Anthropic},
    title = {Model card addendum: Claude 3.5 haiku and upgraded claude 3.5 sonnet},
    year = {2024},
    note = {Technical report, Anthropic},
}

@misc{geminiteam2024gemini15unlockingmultimodal,
      title={Gemini 1.5: Unlocking multimodal understanding across millions of tokens of context}, 
      author={Gemini Team and Petko Georgiev and Ving Ian Lei and Ryan Burnell and Libin Bai and Anmol Gulati and Garrett Tanzer and Damien Vincent and Zhufeng Pan and Shibo Wang and et al},
      year={2024},
      eprint={2403.05530},
      archivePrefix={arXiv},
      primaryClass={cs.CL},
      url={https://arxiv.org/abs/2403.05530}, 
}

@misc{grattafiori2024llama3herdmodels,
      title={The Llama 3 Herd of Models}, 
      author={Aaron Grattafiori and Abhimanyu Dubey and Abhinav Jauhri and Abhinav Pandey and Abhishek Kadian and Ahmad Al-Dahle and Aiesha Letman and Akhil Mathur and Alan Schelten and Alex Vaughan and Amy Yang and Angela Fan and et al},
      year={2024},
      eprint={2407.21783},
      archivePrefix={arXiv},
      primaryClass={cs.AI},
      url={https://arxiv.org/abs/2407.21783}, 
}

@misc{deepseekai2025deepseekv3technicalreport,
      title={DeepSeek-V3 Technical Report}, 
      author={DeepSeek-AI and Aixin Liu and Bei Feng and Bing Xue and et al},
      year={2025},
      eprint={2412.19437},
      archivePrefix={arXiv},
      primaryClass={cs.CL},
      url={https://arxiv.org/abs/2412.19437}, 
}

@inproceedings{reimers-gurevych-2019-sentence,
    title = "Sentence-{BERT}: Sentence Embeddings using {S}iamese {BERT}-Networks",
    author = "Reimers, Nils  and
      Gurevych, Iryna",
    editor = "Inui, Kentaro  and
      Jiang, Jing  and
      Ng, Vincent  and
      Wan, Xiaojun",
    booktitle = "Proceedings of the 2019 Conference on Empirical Methods in Natural Language Processing and the 9th International Joint Conference on Natural Language Processing (EMNLP-IJCNLP)",
    month = nov,
    year = "2019",
    address = "Hong Kong, China",
    publisher = "Association for Computational Linguistics",
    url = "https://aclanthology.org/D19-1410/",
    doi = "10.18653/v1/D19-1410",
    pages = "3982--3992",
}

@inproceedings{NEURIPS2020_3f5ee243,
 author = {Wang, Wenhui and Wei, Furu and Dong, Li and Bao, Hangbo and Yang, Nan and Zhou, Ming},
 booktitle = {Advances in Neural Information Processing Systems},
 editor = {H. Larochelle and M. Ranzato and R. Hadsell and M.F. Balcan and H. Lin},
 pages = {5776--5788},
 publisher = {Curran Associates, Inc.},
 title = {MiniLM: Deep Self-Attention Distillation for Task-Agnostic Compression of Pre-Trained Transformers},
 url = {https://proceedings.neurips.cc/paper_files/paper/2020/file/3f5ee243547dee91fbd053c1c4a845aa-Paper.pdf},
 volume = {33},
 year = {2020}
}

@inproceedings{huang-chang-2023-towards,
    title = "Towards Reasoning in Large Language Models: A Survey",
    author = "Huang, Jie  and
      Chang, Kevin Chen-Chuan",
    editor = "Rogers, Anna  and
      Boyd-Graber, Jordan  and
      Okazaki, Naoaki",
    booktitle = "Findings of the Association for Computational Linguistics: ACL 2023",
    month = jul,
    year = "2023",
    address = "Toronto, Canada",
    publisher = "Association for Computational Linguistics",
    url = "https://aclanthology.org/2023.findings-acl.67/",
    doi = "10.18653/v1/2023.findings-acl.67",
    pages = "1049--1065",
}

@article{williams1992simple,
  title={Simple statistical gradient-following algorithms for connectionist reinforcement learning},
  author={Williams, Ronald J},
  journal={Machine learning},
  volume={8},
  number={3},
  pages={229--256},
  year={1992},
  publisher={Springer}
}

\appendix

\section{Notations}
\label{app::notation}

For clarity and ease of reference, the commonly used notations throughout the paper are summarized in Table~\ref{tab::app-notations}.

\begin{table*}[]
    \centering
    \begin{tabularx}{\textwidth}{cX}
    \hline
        Notation &  Definition \\
    \hline
        $\mathbb{G} = (\mathbb{V}, \mathbb{S}_{edge}, \mathbb{S}_{self}, \mathbb{L})$ &  Candidate space containing roles set $\mathbb{V}$, edge strategies $\mathbb{S}_{edge}$, self-loop strategies $\mathbb{S}_{self}$, LLM pool $\mathbb{L}$ \\
        $\mathbb{V}$ & Set of predefined agent roles (e.g. Analyst, Programmer, Tester) 
        \\
        $\mathbb{S}_{edge}$ & Set of inter-agent collobration strategies (e.g. Debate, Chain) 
        \\
        $\mathbb{S}_{self}$ & Set of internal reasoning strategies (e.g. CoT, Reflection) 
        \\
        $\mathbb{L}$ & Pool of available LLM backbones
        \\
        $\mathcal{G} = (\mathcal{V}, \mathcal{E}, \mathcal{L})$ & Executable graphical multi-agent system
        \\
        $\mathcal{V}$ & Selected roles set for query $q$
        \\
        $\mathcal{E}$ & Selected edges set for query $q$
        \\
        $(u, v, s)$ & The form of the edge, each $(u, v, s) \in \mathcal{E}$
        \\
        $\mathcal{L}$ & Selected llm backbone for each agent $v \in \mathcal{V}$
        \\
        $q$ & Input query to multi-agent system
        \\
        $a$ & Oracle answer corresponding to the query $q$
        \\
        $\mathbb{F}_{\theta} = \mathbb{F}_{\theta_v} \circ \mathbb{F}_{\theta_e} \circ \mathbb{F}_{\theta_l}$ & Controller network for role seletor, edge optimizer, and LLM router
        \\
        $\mathbb{F}_{\theta_v}(\mathcal{V} \mid q)$ & Role seletor, giving the selected roles $\mathcal{V}$ based on query $q$
        \\
        $\mathbb{F}_{\theta_e}(\mathcal{E} \mid q, \mathcal{V})$ & Edge optimizer, giving the selected edges $\mathcal{E}$ based on query $q$ and selected roles  $\mathcal{V}$ 
        \\
        $\mathbb{F}_{\theta_l}(\mathcal{L} \mid q, \mathcal{V}, \mathcal{E})$ & LLM router, giving the selected edges $\mathcal{L}$ based on query $q$, selected roles  $\mathcal{V}$, and selected roles  $\mathcal{E}$ 
        \\
        $\mathbf{H}$ & Latent variable capturing query-roles semantics
        \\
        $\widetilde{\mathbf{H}}_v$ & Refined representation of the candidate agent role $v$
        \\
        $p(a \mid q)$ & Conditional likelihood of generating answer $a$ via MAS
        \\
        $C(\mathcal{G}; q)$ & Cost function quantifying token expenditure 
        \\
        $\lambda$ & Trade-off parameter between utility and cost
        \\
        $\tau$ & Temperature parameter in probability decoding
        \\
        $f_\psi$ & Encoder extracting semantic information from the query $q$
        \\
        $g_\phi$ & Fusion module producing refined representations
        \\
    \hline
    \end{tabularx}
\caption{The notations frequently used throughout this manuscript.}
\label{tab::app-notations}
\end{table*}

\section{Algorithm Workflow}
\label{app::algorithm-workflow}

We conclude the overall algorithm workflow of \textbf{SC-MAS} in Algorithm~\ref{alg::workflow}




\begin{algorithm*}
\renewcommand{\algorithmicrequire}{\textbf{Input:}}
\caption{Workflow of SC-MAS} 
\label{alg::workflow} 
\begin{algorithmic}[1]
\Require Benchmark $\mathcal{D}$, encoder $f_\psi$, fusion module $g_\phi$, learning rate $\alpha$, search space $\mathbb{G} = (\mathbb{V}, \mathbb{S}_{edge}, \mathbb{S}_{self}, \mathbb{L})$
\For{query-answer $(q, a) \in \mathcal{D}$}
    \For{iteration $t \in \{ 1, 2, \cdots, K \}$ }

        \State Sample latent vector $\mathbf{H} \sim \mathcal{N}(\mathbf{H}; \mu_t(q), \text{diag}(\sigma_t^2(q)))$ 
        \State Compute each agent selected probability: $\pi_v(v \mid q, \mathbf{H}) \propto exp(\text{FFN}(f_\psi(q) || \widetilde{\mathbf{H}}_v) / \tau)$
        \State Sampling independently. Selected roles as $\mathcal{V}$

        \State Let $\mathcal{E} \leftarrow$ empty set.
        \For{each selected node $u \in \mathcal{V}$}
            \State Compute self-loop strategies probability: $\pi_{sl}((u, u, s_{sl})$
            \Comment{Eq.(6)}
            \State $\mathcal{E}$ append $(u, u, s_{sl})$ if this strategy is sampled
        \EndFor
        \For{each pair $u, v \in \mathcal{V}$ where $u \ne v$}
            \State Compute edge strategies probability: $\pi_{eg}((u, v, s_{eg})$
            \Comment{Eq.(6)}
            \State $\mathcal{E}$ append $(u, v, s_{eg})$ if this strategy is sampled
        \EndFor

        \State Compute the gnn representations $h_v$ for each $v \in \mathcal{V}$
        \Comment{Eq.(7)}
        \State Compute LLM compatibility: $\pi(l_v = l \mid q, h_v) \propto exp({\text{FFN}(h_v; f_\psi(q))}^\top \mathbf{H}_l / \tau)$
        \State Assign LLM $l_v$ with multinomial samping, symbol as $\mathcal{L}$

        \State Run the MAS $\mathcal{G} = (\mathcal{V}, \mathcal{E}, \mathcal{L})$ 
        \Comment{Algorithm (1)}
        
        \State Compute reward $R = U(\mathcal{G}; q, a) - \lambda \cdot C(\mathcal{G}, q)$
        \Comment{Eq.(9)}
        \State Update $\theta$ via policy gradient: $\theta \leftarrow \theta - \alpha  \nabla_\theta \mathbf{E}[-R]$
        
    \EndFor
\EndFor
\end{algorithmic} 
\end{algorithm*}

\section{Detailed Cost-Performance Data}
\label{app::cost-any}

\subsection{Inference Cost}
In this section, we provide a detailed assessment of both the computational overhead and the performance of selected baselines on the MBPP and HumanEval datasets, as summarized in Table~\ref{tab::inference-mbpp-humaneval}.

\begin{table*}[htbp]
\centering
\begin{tabular}{c|c|cc|cc}
\hline
\multirow{2}{*}{\textbf{Method}}         & \multirow{2}{*}{\textbf{LLM}} & \multicolumn{2}{c|}{\textbf{MBPP}} & \multicolumn{2}{c}{\textbf{HumanEval}} \\
                                &                      & Score(\%)    & Cost(\$)   & Score(\%)      & Cost(\$)     \\ \hline
\multirow{4}{*}{IO}             & gpt-4o-mini          & 72.20        & 0.143      & 85.71          & 0.025        \\
                                & claude-3.5-haiku     & 73.40        & 0.146      & 86.33          & 0.025        \\
                                & gemini-1.5-flash     & 73.00        & 0.157      & 82.61          & 0.032        \\
                                & llama-3.1-70b        & 68.20        & 0.105      & 80.75          & 0.013        \\
\multirow{2}{*}{Complete Graph} & gpt-4o-mini          & 75.20        & 3.088      & 85.00          & 0.488        \\
                                & gemini-1.5-flash     & 74.20        & 1.215      & 83.75          & 0.568        \\
\multirow{2}{*}{LLM Debate}     & gpt-4o-mini          & 73.60        & 4.427      & 84.38          & 0.624        \\
                                & gemini-1.5-flash     & 73.40        & 4.529      & 79.38          & 0.693        \\
\multirow{2}{*}{AgentPrune}     & gpt-4o-mini          & 75.00        & 1.215      & 86.80          & 0.254        \\
                                & gemini-1.5-flash     & 75.60        & 1.352      & 82.55          & 0.271        \\
\multirow{2}{*}{AFlow}          & gpt-4o-mini          & 82.20        & 1.723      & 90.15          & 0.363        \\
                                & gemini-1.5-flash     & 76.00        & 1.832      & 85.69          & 0.386        \\
FrugalGPT                       & llm pool             & 74.40        & 0.139      & 87.31          & 0.026        \\
RouterDC                        & llm pool             & 75.20        & 0.145      & 87.75          & 0.023        \\
MasRouter                       & llm pool             & 84.00        & 1.039      & 90.52          & 0.185        \\
SC-MAS                      & llm pool             & 87.53        & 0.913      & 92.37          & 0.154        \\ \hline
\end{tabular}
\caption{Inference Cost-Performance on MBPP and HumanEval Dataset.}
\label{tab::inference-mbpp-humaneval}
\end{table*}

\subsection{Traing Cost}
\label{sec::app-traing-cost}

We employed an 8G GPU for training, though most of the time was spent waiting for LLM API responses, which depend on the service provider, making direct time comparisons less meaningful. 
As shown in Table~\ref{tab::train-cost}, we evaluated the training overhead on the MATH and MMLU datasets in comparison to other state-of-the-art methods such as GPTSwarm \cite{gptswarm2024}, AFlow \cite{zhang2025aflow}, MasRouter \cite{yue2025masrouter}, and our results indicated a substantial reduction in token consumption. Due to our representation of the collaboration strategy as edges, additional time is required to learn the strategy’s relational structures, resulting in higher training overhead than MasRouter. Nevertheless, our completion token usage remains lower than that of MasRouter, highlighting the benefits of a heterogeneous collaboration strategy.


\begin{table*}[htbp]
\begin{tabular*}{\hsize}{@{}@{\extracolsep{\fill}}ccccccc@{}}
\hline
\multirow{2}{*}{\textbf{Method}} & \multicolumn{3}{c}{\textbf{MATH}}                                                                                                                                                          & \multicolumn{3}{c}{\textbf{MMLU}}                                                                                                                                                          \\ \cline{2-7} 
                        & \begin{tabular}[c]{@{}c@{}}Prompt\\ tokens\end{tabular} & \begin{tabular}[c]{@{}c@{}}Completion\\ tokens\end{tabular} & \begin{tabular}[c]{@{}c@{}}Total\\ cost (\$)\end{tabular} & \begin{tabular}[c]{@{}c@{}}Prompt\\ tokens\end{tabular} & \begin{tabular}[c]{@{}c@{}}Completion\\ tokens\end{tabular} & \begin{tabular}[c]{@{}c@{}}Total\\ cost (\$)\end{tabular} \\ \hline
GPTSwarm   & 23,031,287   & 6,943,173   & 7.63   & 15,525,155 & 3,983,745  & 4.70   \\
AFlow      & 321,813,314  & 28,083,445  & 21.75  & 13,085,019 & 11,239,502 & 8.67   \\
MasRouter  & 3,235,288    & 2,499,530   & 3.56   & 4,459,674  & 2,904,656  & 1.43   \\
SC-MAS & 17,672,292   & 1,852,574   & 5.08   & 12,808,047  & 2,083,184  & 2.64  \\ 
\hline
\end{tabular*}
\caption{Training Cost comparison between SC-MAS and state-of-the-art baselines on MATH and MMLU.}
\centering
\label{tab::train-cost}
\end{table*}

\section{Case Study}
\label{app::case-study}

As shown in Tables \ref{case-study::mbpp} to \ref{case-study::mmlu}, we visualize the customized MAS designed by SC-MAS for varying query difficulties on the five benchmark.
In this graph, blue nodes represent the start nodes and orange nodes mark the output nodes; the entire MAS graph is executed according to Algorithm~\ref{alg1}, ultimately generating the final output.
In this representation of heterogeneous collaboration strategies, two agents automatically establish appropriate collaboration mode based on their relationship with each other.

\begin{table*}[htbp]
    \centering
    \begin{tabular*}{\hsize}{@{}@{\extracolsep{\fill}}c|c@{}} 
\hline 
\textbf{Query} 
& 
\textbf{SC-MAS Workflow} 
\\ \hline
\makecell[l]{
Write a python function to find the minimum \\ number of rotations (greater than 0) required \\ to get the same string.  \\
\\
You need to pass these test cases: \\
\\
assert find\_Rotations("aaaa") == 1 \\
assert find\_Rotations("ab") == 2 \\
assert find\_Rotations("abc") == 3 \\
}
& 
        \begin{minipage}[b]{0.98\columnwidth}
		\centering
		\raisebox{-.5\height}{\includegraphics[width=\linewidth]{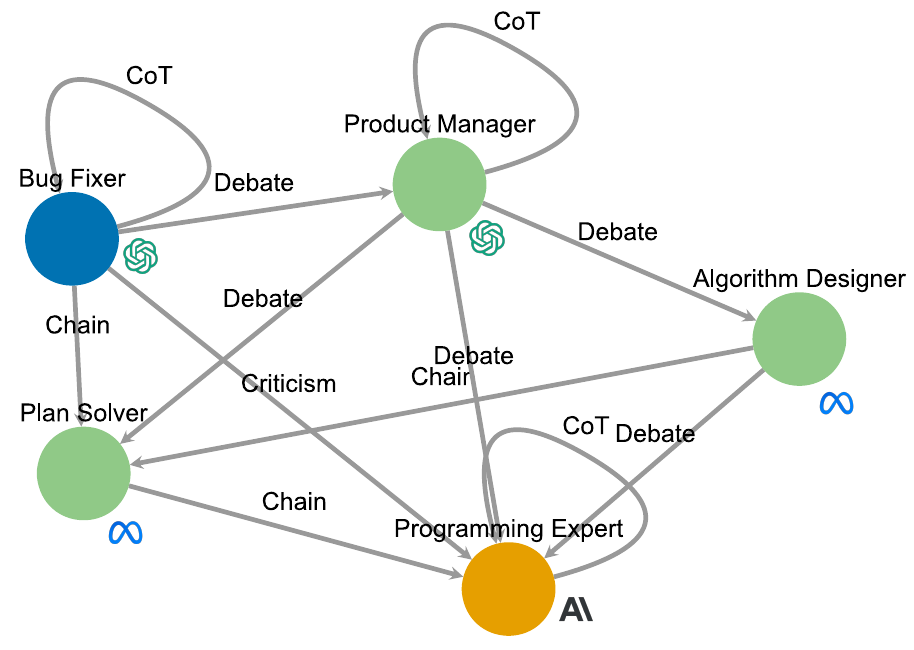}}
	\end{minipage}
\\ \hline
\makecell[l]{
Write a function to flatten a list and sum \\ all of its elements. \\
\\ 
You need to pass these test cases: \\
\\
assert recursive\_list\_sum( \\
\hspace{2em} ([1, 2, [3,4],[5,6]]))==21 \\
assert recursive\_list\_sum( \\
\hspace{2em} ([7, 10, [15,14],[19,41]]))==106 \\
assert recursive\_list\_sum( \\
\hspace{2em} ([10, 20, [30,40],[50,60]]))==210 \\
}
&  
        \begin{minipage}[b]{0.98\columnwidth}
		\centering
		\raisebox{-.5\height}{\includegraphics[width=\linewidth]{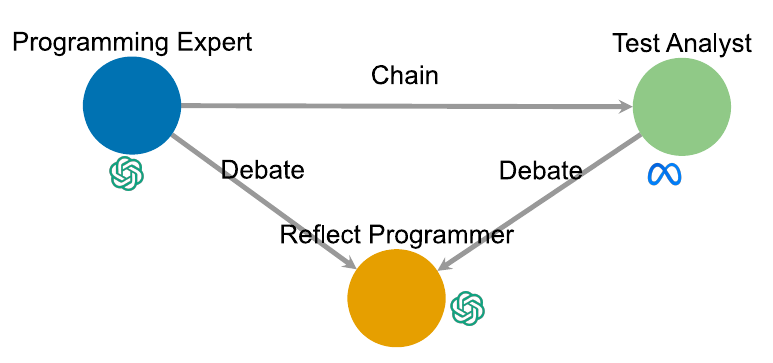}}
	\end{minipage}
\\ \hline
    \end{tabular*}
\caption{Case study on MBPP dataset.}
\label{case-study::mbpp}
\end{table*}

\begin{table*}[htbp]
    \centering
    \begin{tabular*}{\hsize}{@{}@{\extracolsep{\fill}}c|c@{}}
\hline 
\textbf{Query} 
& 
\textbf{SC-MAS Workflow} 
\\ \hline
\makecell[l]{
A fruit vendor bought 50 watermelons for \\ \$80. He sold all of them at a profit of 25\%. \\ How much was each watermelon sold?
}
& 
        \begin{minipage}[b]{0.98\columnwidth}
		\centering
		\raisebox{-.5\height}{\includegraphics[width=\linewidth]{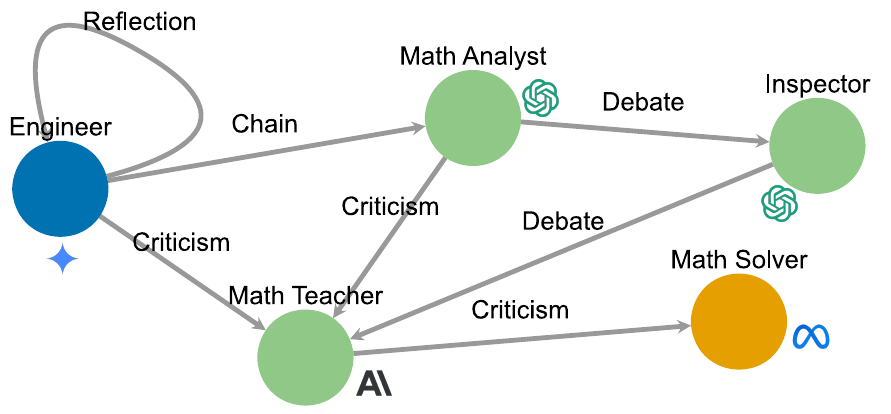}}
	\end{minipage}
\\ \hline
\makecell[l]{
A restaurant has 40 tables with 4 legs and \\ 50 tables with 3 legs. Calculate the total \\ number of legs the restaurant's tables have.
}
&  
        \begin{minipage}[b]{0.98\columnwidth}
		\centering
		\raisebox{-.5\height}{\includegraphics[width=\linewidth]{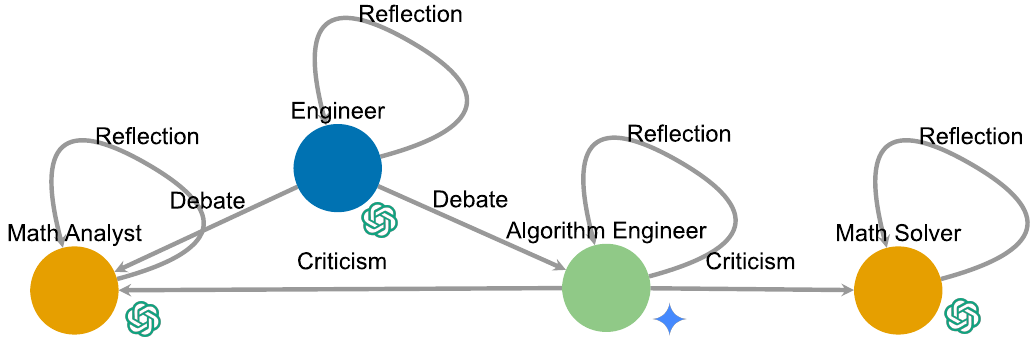}}
	\end{minipage}
\\ \hline
    \end{tabular*}
\caption{Case study on GSM8K dataset.}
    \label{case-study::gsm8k}
\end{table*}

\begin{table*}[htbp]
    \centering
    \begin{tabular*}{\hsize}{@{}@{\extracolsep{\fill}}c|c@{}} 
\hline 
\textbf{Query} 
& 
\textbf{SC-MAS Workflow} 
\\ \hline
        \makecell[l]{
def is\_prime(n): \\
\hspace{2em} """Return true if a given number \\ is prime, and false otherwise. \\
\hspace{2em}    >> is\_prime(6) \\
\hspace{2em}    False \\
\hspace{2em}    >> is\_prime(101) \\
\hspace{2em}    True \\
\hspace{2em}    >> is\_prime(11) \\
\hspace{2em}    True \\
\hspace{2em}    >> is\_prime(13441) \\
\hspace{2em}    True \\
\hspace{2em}    >> is\_prime(61) \\
\hspace{2em}    True \\
\hspace{2em}    >> is\_prime(4) \\
\hspace{2em}    False \\
\hspace{2em}    >> is\_prime(1) \\
\hspace{2em}    False \\
\hspace{2em}    """
}
& 
        \begin{minipage}[b]{0.98\columnwidth}
		\centering
		\raisebox{-.5\height}{\includegraphics[width=\linewidth]{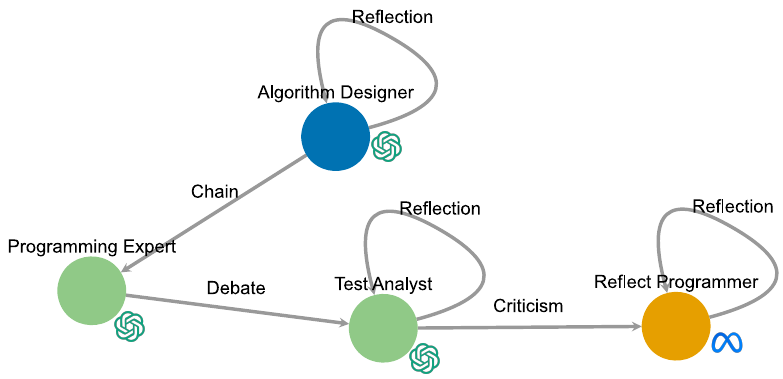}}
	\end{minipage}
\\ \hline
\makecell[l]{
def find\_closest\_elements(numbers: List[ \\ 
\hspace{2em} float]) -> Tuple[float, float]: \\
\hspace{2em}    """ From a supplied list of numbers \\
    (of length at least two) select and return \\ 
    two that are the closest to each other and   \\ 
    return them in order (smaller number,   \\
    larger number). \\
\hspace{2em}    >> find\_closest\_elements( \\ 
\hspace{4em} [1.0, 2.0, 3.0, 4.0, 5.0, 2.2]) \\
\hspace{2em}    (2.0, 2.2) \\
\hspace{2em}    >> find\_closest\_elements( \\ 
\hspace{4em} [1.0, 2.0, 3.0, 4.0, 5.0, 2.0]) \\
\hspace{2em}   (2.0, 2.0) \\
\hspace{2em}   """
}
&  
        \begin{minipage}[b]{0.98\columnwidth}
		\centering
		\raisebox{-.5\height}{\includegraphics[width=\linewidth]{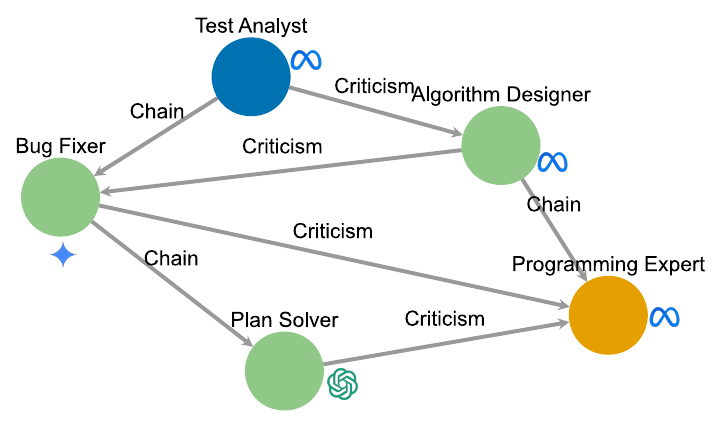}}
	\end{minipage}
\\ \hline
\makecell[l]{
def smallest\_change(arr): \\
\hspace{2em}    """ \\
\hspace{2em}    Given an array arr of integers, find \\ the minimum number of elements that \\ need to be changed to make the array \\ palindromic. A palindromic array is an \\ array  that is read the same backwards \\ and forwards. In one change, you can \\ change one element to any other element.\\
\\
    For example: \\
    smallest\_change([1,2,3,5,4,7,9,6]) == 4 \\
    smallest\_change([1, 2, 3, 4, 3, 2, 2]) == 1 \\
    smallest\_change([1, 2, 3, 2, 1]) == 0 \\
    """
}
&  
        \begin{minipage}[b]{0.98\columnwidth}
		\centering
		\raisebox{-.5\height}{\includegraphics[width=\linewidth]{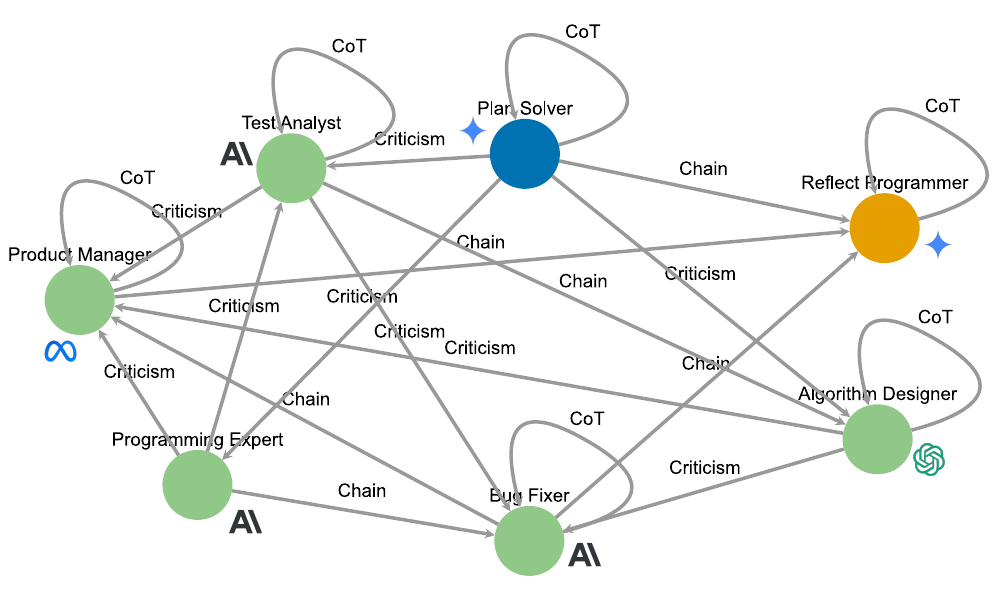}}
	\end{minipage}
\\ \hline
    \end{tabular*}
\caption{Case study on HumanEval dataset.}
    \label{case-study::humaneval}
\end{table*}

\begin{table*}[htbp]
    \centering
    \begin{tabular*}{\hsize}{@{}@{\extracolsep{\fill}}c|c@{}} 
\hline 
\textbf{Query} 
& 
\textbf{SC-MAS Workflow} 
\\ \hline
\makecell[l]{
What is the positive difference \\ between $120\%$ of 30 and $130\%$ of 20?
}
& 
        \begin{minipage}[b]{0.85\columnwidth}
		\centering
		\raisebox{-.5\height}{\includegraphics[width=\linewidth]{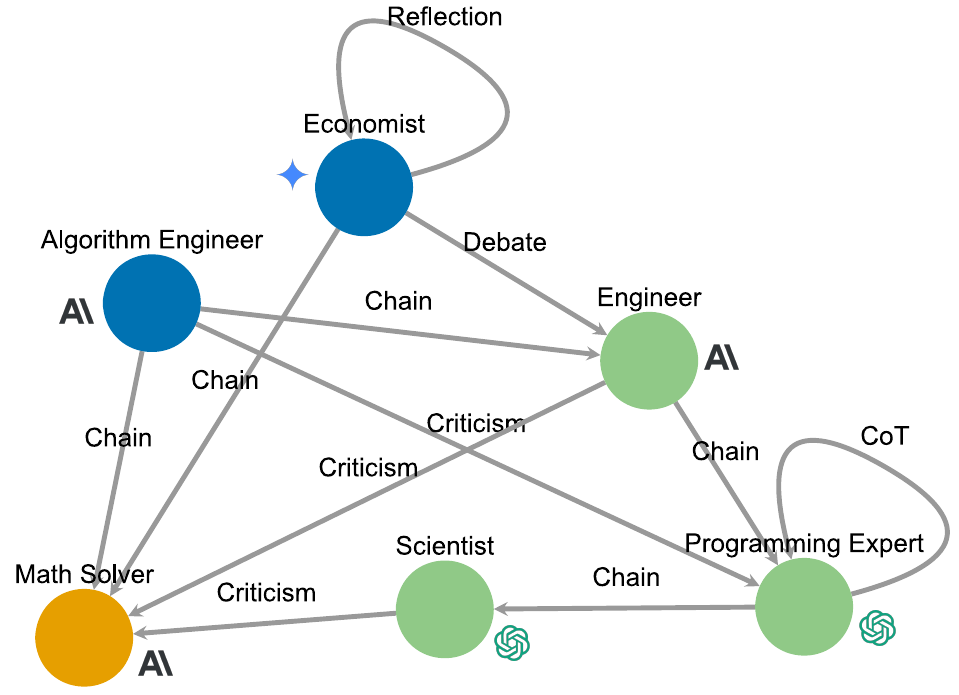}}
	\end{minipage}
\\ \hline
\makecell[l]{
In $\triangle{ABC}$, shown, $\cos{B}=\frac{3}{5}$. \\ 
What is $\cos{C}$? \\
\\
>> asy \\
draw((0,0)--(4,0)--(0,3)--\\ 
\hspace{2em} cycle,black+linewidth(1)); \\
draw(rightanglemark((4,0),(0,0),(0,3),10),\\ 
\hspace{2em} black+linewidth(1)); \\
label("$A$",(0,0),W); \\
label("$B$",(0,3),W); \\
label("$C$",(4,0),E); \\
label("9",(0,0)-\-(0,3),W); \\
>> asy
}
&  
        \begin{minipage}[b]{0.85\columnwidth}
		\centering
		\raisebox{-.5\height}{\includegraphics[width=\linewidth]{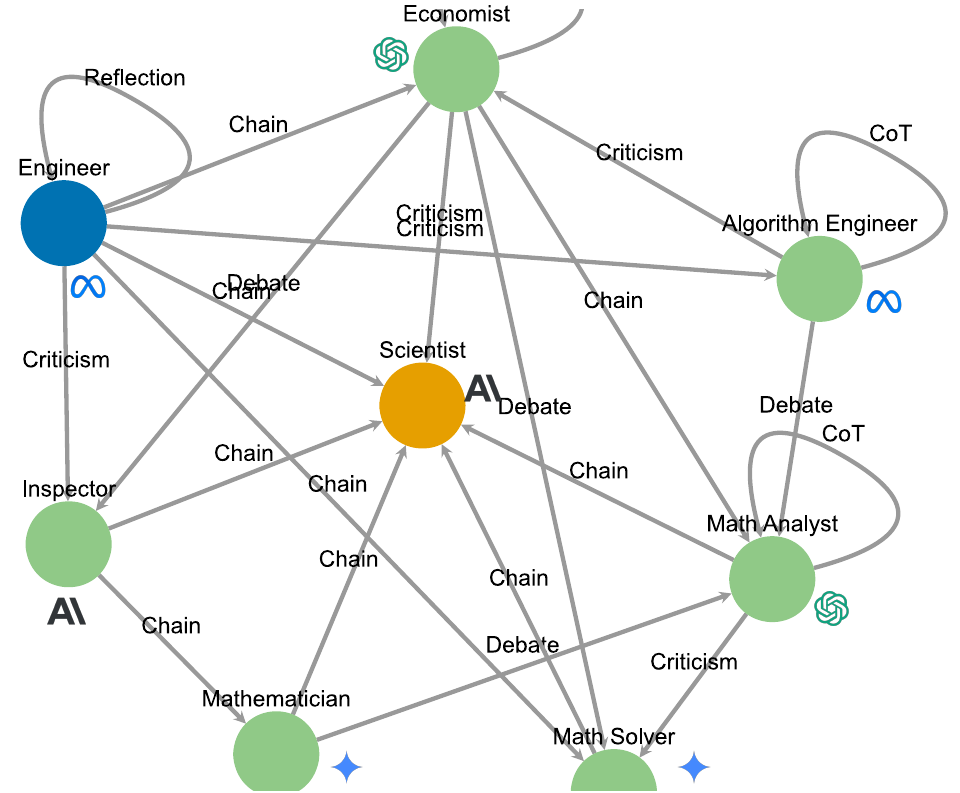}}
	\end{minipage}
\\ \hline
    \end{tabular*}
\caption{Case study on MATH dataset.}
    \label{case-study::math}
\end{table*}

\begin{table*}[htbp]
    \centering
    \begin{tabular*}{\hsize}{@{}@{\extracolsep{\fill}}c|c@{}} 
\hline 
\textbf{Query} 
& 
\textbf{SC-MAS Workflow} 
\\ \hline
\makecell[l]{
A comet of mass m impacts the earth \\ (mass M radius R) at the minimum impact \\ speed. What is the expression for the total \\ energy released in the impact?, \\ $m*v,0.5*m/(R^3),0.5*m*(2GM/R)$ \\
$,0.6*G(M^2)/R,C$
}
& 
        \begin{minipage}[b]{0.85\columnwidth}
		\centering
		\raisebox{-.5\height}{\includegraphics[width=\linewidth]{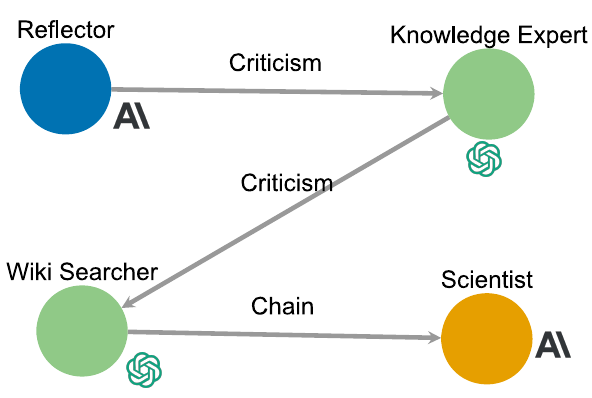}}
	\end{minipage}
\\ \hline
\makecell[l]{
Which of the following gives the total \\ spin quantum number of the electrons in \\ the ground state of neutral nitrogen  \\ $(Z = 7)?,1/2,1,3/2,5/2,C$
}
&  
        \begin{minipage}[b]{0.85\columnwidth}
		\centering
		\raisebox{-.5\height}{\includegraphics[width=\linewidth]{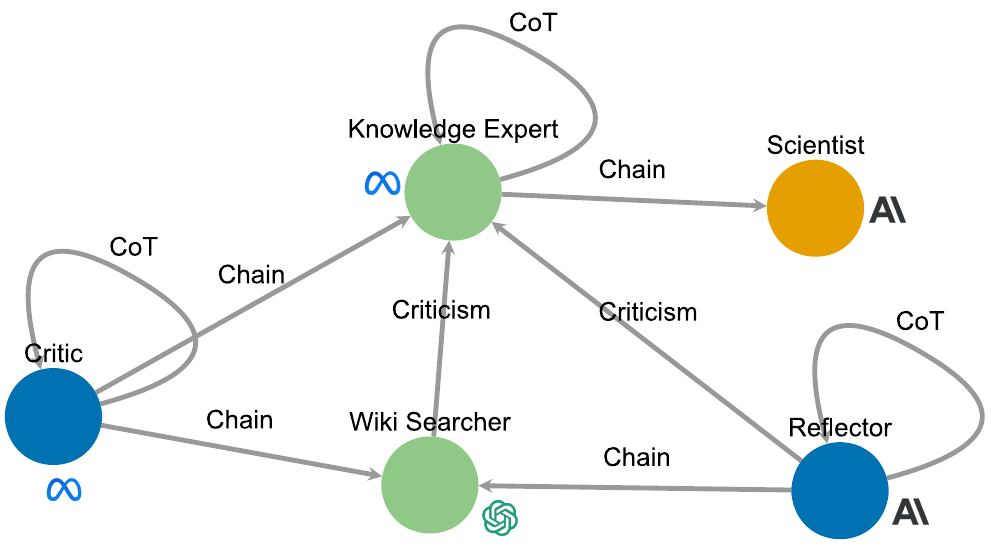}}
	\end{minipage}
\\ \hline
    \end{tabular*}
\caption{Case study on MMLU dataset.}
    \label{case-study::mmlu}
\end{table*}

\end{document}